\documentclass[%
 aip,
 amsmath,amssymb,
 reprint,%
]{revtex4-1}

\usepackage{graphicx}
\usepackage{dcolumn}
\usepackage{bm}
\usepackage[table,dvipsnames]{xcolor}
\usepackage{tikz}
\usepgflibrary{shapes.geometric}
\usetikzlibrary{arrows, decorations.pathmorphing, patterns}
\usepackage[utf8]{inputenc}
\usepackage[T1]{fontenc}
\usepackage{mathptmx}
\usepackage{etoolbox}
\usepackage{braket}
\usepackage[dvipsnames]{xcolor}

\usepackage[mathlines]{lineno}

\makeatletter
\def\@email#1#2{%
 \endgroup
 \patchcmd{\titleblock@produce}
  {\frontmatter@RRAPformat}
  {\frontmatter@RRAPformat{\produce@RRAP{*#1\href{mailto:#2}{#2}}}\frontmatter@RRAPformat}
  {}{}
}%
\makeatother
\begin{document}

\preprint{AIP/123-QED}

\title[Sample title]{An effective bath state approach to model infrared spectroscopy and intramolecular dynamics in complex molecules}
\author{Loïse Attal}
\affiliation{Université Paris-Saclay, CNRS, Institut des Sciences Moléculaires d'Orsay, 91405 Orsay, France
}%
\affiliation{Theoretische Chemie, Institut für Chemie, Universität Potsdam, D-14476 Potsdam-Golm, Germany}
\email{loise.attal@uni-potsdam.de}
 
\author{Cyril Falvo}%
\affiliation{Université Paris-Saclay, CNRS, Institut des Sciences Moléculaires d'Orsay, 91405 Orsay, France
}%
\affiliation{%
	Université Grenoble-Alpes, CNRS, LIPhy, 38000 Grenoble, France
}%

\author{Pascal Parneix}
\affiliation{Université Paris-Saclay, CNRS, Institut des Sciences Moléculaires d'Orsay, 91405 Orsay, France
}%


\begin{abstract}
When a molecule contains more than a few atoms, its full-dimensional dynamics becomes untractable, especially when introducing temperature effects. In such a case, it can be interesting to focus only on a few degrees of freedom and to model the rest of the molecule as a finite-dimensional bath. 
In this prospect, we extend the effective bath state (EBS) method that we had first developed and benchmarked in [J. Chem. Phys. \textbf{160}, 044107 (2024)] to describe the spectroscopy and intramolecular dynamics of complex isolated molecules. 
The EBS method is a system-bath approach based on the coarse-graining of the bath into a reduced set of effective energy states. It allows for a significant reduction of the bath dimension and makes finite-temperature calculations more accessible. In order to treat a realistic molecule, the method is extended to include polynomial couplings in the bath coordinates. The ability of the method to model temperature-resolved infrared spectra and to follow population transfers between the vibrational modes of the molecule is first tested on a 10-mode model system. The extended method is then applied to the realistic case of phenylacetylene.
\end{abstract}

\maketitle

\section{Introduction}

Quantum dynamics provides the most complete description of the evolution of molecular systems by accounting for their quantized nature and for purely quantum processes such as delocalization or tunneling. However, the effort to solve the underlying Schrödinger equation grows exponentially with the number of degrees of freedom (DOFs), making it virtually impossible to treat complex systems exactly. This exponential wall, the so-called curse of dimensionality,\cite{Donoho_curse_2000} is particularly striking for systems in contact with an environment, such as molecules or atoms on surfaces\cite{Morse_Surface_2012,Bonfanti_2015_jcpI,Bonfanti_2015_jcpII,bouakline_quantum-mechanical_2019}, charge transfers in liquids \cite{Yan_Solvation_1988} or solids \cite{wang_semiclassical_1999,Craig_2007_jcp}, or qubits in contact with a solid-state environment.\cite{Benenti_Emergence_2001,halbertal_nanoscale_2016,Pekola_finite-size_2016} But this dimensionality issue can also occur inside large isolated molecules, where the dimension of the vibrational problem ($g =3N -6$ for a molecule with $N$ atoms) can quickly cause full-dimensional methods to be untractable. 
It also affects the calculation of infrared (IR) absorption spectra, since they are usually obtained from the Fourier transform of the dipole autocorrelation functions\cite{Tuckerman_book_2001,Breuer:2007ab} that are obtained from quantum dynamical calculations. 

The difficulties mentioned above are amplified when including temperature effects. Indeed, at  $T = 0$~K, the only possible initial state is the ground state. Whereas, for non-zero temperatures, many initial states can be populated and have to be taken into account. A finite-temperature spectrum or dynamics results from a large number of energy-fixed trajectories that are averaged using the corresponding Boltzmann distribution. This makes temperature effects computationally expensive, and full-dimensional methods are usually restricted to 0~K calculations.\cite{manthe_wavepacket_1992,Nest_Meyer_2003,Morse_Surface_2012,Schroder2019,Xie_2019,fischer_hierarchical_2020} 
A new scheme has recently opened a very promising path to overcome this issue, by reproducing temperature effects a posteriori from a single 0~K wavefunction simulation.\cite{tamascelli_efficient_2019,Dunnett_2021_fchem} Apart from this work, semi-classical approaches are usually the preferred solution to introduce temperature effects in the dynamics of the nuclei.\cite{Bridge_Quantum_2024,Makri_Parsing_2024,Lanzi_semiclassical_2025}

For systems where full-dimensional treatment is not feasible, a common alternative is to turn to system-bath methods where the problem is divided into a system  (e.g., an atom) and an environment (e.g., the surface on which it is adsorbed), which is often called a bath. The system is treated as rigorously as possible, but numerical gains are made on the bath, either by tracing it out completely,\cite{REDFIELD1965,Breuer:2001aa,Breuer:2007ab,Tanimura_review_2020_jcp,Dunnett_2021_fchem,Finite_bath_2022} or by strongly simplifying its description,\cite{Baer_Quantum_1997,fischer_hierarchical_2020} reducing its effective dimension,\cite{fischer_hierarchical_2020} or introducing semiclassical approximations.\cite{Mukamel_Femtosecond_1990,Fried_chapter_1993}

System-bath approaches can be applied naturally to small systems in contact with a macroscopic environment (like a surface or a liquid), but a large isolated molecule can also require a system-bath treatment. In that case, some of the molecule DOFs are selected to be the system of interest, and the bath is made of the remaining ones. However, in such an intra-molecular context, an additional difficulty arises: the bath is composed of a finite number of DOFs. Because of that, it cannot be described by usual open quantum system methods, which assume the bath to be infinite and trace it out completely.\cite{Scully_quantum_1967,MOLLOW1969464,Breuer:2007ab,Tanimura_review_2020_jcp} By definition, an infinite bath is always at thermodynamical equilibrium and is not influenced by the evolution of the system. This is not true inside a molecule: the excitation of a given vibrational mode will have an impact on the evolution of the rest of the molecule.

Even though system-bath methods have started from the Markovian and perturbative approximations,\cite{REDFIELD1965,Linblad_1976} many methods are now able to go beyond these approximations by accounting for memory effects,\cite{Tanimura_HEOM_1989,Gaspard_non_Markov_1999,Breuer:2001aa, Vega_non-Markovian_2017,Tamascelli_Nonperturbative_2018, Tanimura_review_2020_jcp,Teixeira_weak-coupling_2022} or reaching the strong coupling regime.\cite{Prior_Efficient_2010, Tamascelli_Nonperturbative_2018,Mandal_SB_strong_2022}  
Some authors have also introduced several levels of baths with different levels of approximations to reach larger environments while keeping a more rigorous description of the main bath modes.\cite{Burghardt_Approaches_1999,Picconi_high-dimensional_2024,Mandal_Quantized_2025}

However, all the aforementioned strategies were conceived in the context of infinite baths, and the situation of an intermediate-sized bath, too large to be treated exactly but too small to be unperturbed by the main system, has not been extensively studied so far, except in the work of Esposito and Gaspard.\cite{esposito_quantum_2003,esposito_spin_2003,esposito_quantum_2007} In their contributions, the authors introduce a microcanonical master equation that takes into account the influence of the system's evolution on the state of a finite bath, together with rigorous conservation of the total energy. A similar approach, called the extended microcanonical master equation, was recently developed and used to describe the non-equilibrium dynamics of the central spin method.\cite{Riera-Campeny:2021aa,Finite_bath_2022}  

In that context, we have developed the effective bath state (EBS) method,\cite{Attal_EBS_model_2024} a wavefunction-based system-bath approach where the studied system is seen as a one-dimensional sub-system coupled to a finite bath. When studying an isolated molecule with vibrational modes $g$, the EBS method is applied by taking one mode of interest as the system and considering the remaining $g-1$ modes as a finite-size environment surrounding this system. The mode of interest is treated rigorously by considering its intra-mode anharmonicity and its coupling to the other modes. The bath description is simplified by assuming that all bath modes are harmonic and uncoupled. 
The central idea of the EBS method is to transform the $g-1$ bath modes into a single ladder of effective energy states (EESs) representing the total bath energy. The use of EESs allows for a significant reduction of the bath dimension and of its scaling with the number of bath modes. 
Contrary to usual open quantum systems methods, the EBS method keeps some information about the state of the bath through the EESs. Together with realistic system-bath couplings obtained from quantum chemistry calculations, this allows us to follow the time evolution of both the system and the bath in realistic situations. With this (semi-)explicit representation of the bath, the EBS method can account for non-Markovian system-bath interactions and reproduce effects due to the finite size of the bath.
Defining global bath energy states also enables the direct preparation of the bath at a given, possibly non-zero, energy. This facilitates the simulation of finite temperature effects in relatively large molecules by avoiding an expensive sampling of the initial states. 

In a previous paper,\cite{Attal_EBS_model_2024} we have introduced the EBS formalism and tested its capacities in the case of a vibrational stretching mode interacting with a bath of 40 harmonic oscillators. 
This model situation was taken from Ref.~\onlinecite{Morse_Surface_2012} and an excellent agreement was found between our calculations and the results obtained by the authors using the multiconfigurational time-dependent Hartree (MCTDH) method. It was also shown that the EBS method could go beyond these results by considering larger baths (up to 600 modes) and by including finite-energy or finite-temperature effects.\cite{Attal_EBS_model_2024} However, only linear coupling terms in the bath coordinates were used in this model system\cite{Morse_Surface_2012} and no spectral quantities were computed. Here, we extend the EBS method by including polynomial couplings in the bath coordinates in order to treat realistic molecular systems. This is an essential point, since most system-bath or effective mode methods can only deal with linear couplings in the bath coordinates.\cite{CALDEIRA1981, NEST200165,fischer_hierarchical_2020,Tanimura_review_2020_jcp} We also extend the method to compute IR absorption spectra at finite temperatures.

The article is organized as follows: the derivation of the EBS method and its most recent extensions are detailed in Sec.~\ref{sec:method}. The extended method is then tested on a 10-mode model system (Sec.~\ref{sec:Spectra_10modes}) before being applied to a realistic system, namely phenylacetylene, in Sec.~\ref{sec:Ph-Ace}. Finally, some conclusions and perspectives are given in Sec.~\ref{sec:conclusion}.

\section{The EBS method in a molecular context}
\label{sec:method}

\subsection{Hamiltonian}
\label{subsec:S-B_Hamiltonian}

We consider a $g$-dimensional molecular system in a given electronic state, described by its normal coordinates $\hat{\bf{Q}} = \{\hat{Q}_1,\hdots, \hat{Q}_g\}$ and their associated momenta $\hat{\bf{P}} = \{\hat{P}_1,\hdots, \hat{P}_g\}$. We expand the potential energy surface (PES) as a quartic polynomial in the normal mode coordinates, leading to the following Hamiltonian: 

\begin{equation}
\begin{aligned}
\hat{H}({\bf{\hat{Q}}},{\bf{\hat{P}}}) 
& = \sum_i \frac{\hat{P}_i^2}{2} + \frac{1}{2} \omega_i^2 \, \hat{Q}_i^2 + \frac{1}{3!} \sum_{i,j,k} \alpha_{ijk} \,\hat{Q}_i\hat{Q}_j\hat{Q}_k  \\
&+ \frac{1}{4!}  \sum_{i,j,k,l} \beta_{ijkl} \, \hat{Q}_i\hat{Q}_j\hat{Q}_k\hat{Q}_l \,.
\label{eq:quartic_PES}
\end{aligned}
\end{equation}
Coupling coefficients $\alpha_{ijk}$ and $\beta_{ijkl}$ can be obtained from quantum chemistry calculations, or they can be model parameters as is the case for Ohmic baths.\cite{Nest_Meyer_2003, Morse_Surface_2012}

To obtain a system-bath Hamiltonian, the molecule is divided into a mode of interest that will be seen as the system, and a bath containing all the other normal modes. Without loss of generality, the mode of interest will be denoted as mode~1. 
Following standard system-bath conventions, the Hamiltonian $\hat{H}$ is divided into three parts associated with the system ($\hat{H}_{\rm{S}}$), the bath ($\hat{H}_{\rm{B}}$), and the system-bath interaction ($\hat{H}_{\rm{SB}}$), respectively:

\begin{equation}
\hat{H}(\hat{\bf{Q}},\hat{\bf{P}}) = \hat{H}_{\rm{S}}(\hat{Q}_1,\hat{P}_1) + \hat{H}_{\rm{B}}(\hat{\bf{Q}}_{i \ne 1},\hat{\bf{P}}_{i \ne 1}) +\hat{H}_{\rm{SB}}(\hat{\bf{Q}})\, .
\label{eq:H_tot}
\end{equation}

\subsubsection{System Hamiltonian}
The system Hamiltonian is composed of all the terms that only relate to the mode of interest, and corresponds to the intra-mode anharmonic Hamiltonian of mode 1:

\begin{equation}
\hat{H}_{\rm S} = \frac{\hat{P}_{1}^2}{2} + \frac{1}{2} \omega_{1}^2 \, \hat{Q}_{1}^2 + \frac{1}{3!} \alpha_{111} \,\hat{Q}_{1}^3+ \frac{1}{4!}   \beta_{1111} \, \hat{Q}_{1}^4\, .
\label{eq:H_S_quartic}
\end{equation}
The EBS method makes no further approximations on the description of the system, and the above Hamiltonian can be diagonalized to obtain

\begin{equation}
    \hat{H}_{\rm S} = \sum_{v= 0}^{N_v -1} E_v \ket{v}\bra{v}\,,
\end{equation}
where $E_v$ are the system eigenenergies, $\ket{v}$ its eigenstates, and $N_v$ is the number of system eigenstates considered in a given numerical calculation.

Note that, since we have access to the eigenstates of the system, $\hat{H}_{\rm S}$ is actually not restricted to polynomial functions. It can be any function of the system coordinate. Hence, any Hamiltonian of the form
\begin{equation}
    \hat{H}_{\rm S} = \frac{\hat{P}_{1}^2}{2} + \hat{V}_{\rm S}(\hat{Q}_1),
\end{equation}
with $\hat{V}_{\rm S}$ a (reasonably smooth) potential operator, can be used to describe the system. This notably allows the mode of interest to be represented by a Morse potential,\cite{Morse_1929} as was the case in our previous article.\cite{Attal_EBS_model_2024}

\subsubsection{Bath Hamiltonian}
The bath is made of the $g-1$ remaining normal modes ($i = 2,\hdots, g$). The Hamiltonian of Eq.~\eqref{eq:quartic_PES} contains both a harmonic part and anharmonic coupling terms. However, in the EBS method the internal bath couplings are neglected and the bath modes are assumed to be harmonic. The bath Hamiltonian is thus taken as

\begin{equation}
    \hat{H}_{\rm B} =\sum_{i = 2}^{g} \frac{\hat{P}_i^2}{2} + \frac{1}{2} \omega_i^2 \, \hat{Q}_i^2\, .
    \label{eq:H_bath_harmo}
\end{equation}
The Hamiltonian above can be written in the harmonic basis set $\ket{\bf n}=\ket{n_2, n_3,  ..., n_g}$ as

\begin{equation}
\hat{H}_{\rm{B}}
=\sum_{{\bf n}} E({\bf n}) \ket{{\bf n}}\bra{{\bf n}}\, ,
\label{eq:bath_exact}
\end{equation}
with  $E({\bf n}) = \sum_{i = 2}^g n_i\hbar \omega_i$. In the following, the harmonic microstates $\ket{\bf n}$ will be used as reference states for the bath.

\subsubsection{System-bath coupling Hamiltonian}
\label{subsec:def_H_SB}
The system-bath coupling Hamiltonian regroups all the terms connecting the system (mode 1) to one or several bath modes. 
In the following, we only keep the terms that give a non-vanishing contribution within second-order vibrational perturbation theory,\cite{Dunham_coeffs,Sibert_PT_1988,McCoy_rotation-vibration_1992} and consider the following coupling Hamiltonian: 
\begin{equation}
\begin{aligned} 
\hat{H}_{\rm SB}&=  \frac{1}{2}\sum_{k> 1} \alpha_{11k}  \,\hat{Q}_{1}^2\hat{Q}_k + \frac{1}{2} \sum_{k > 1} \alpha_{1kk} \,\hat{Q}_{1}\hat{Q}_k^2 \\ 
&  + \sum_{k> j>1} \alpha_{1jk} \,\hat{Q}_{1}\hat{Q}_j\hat{Q}_k + \frac{1}{4}  \sum_{k > 1} \beta_{11kk} \, \hat{Q}_{1}^2\hat{Q}_k^2  \, .
\end{aligned}
\label{eq:H_coupling_total}
\end{equation}

For the same reasons as for $\hat{H}_{\rm S}$, the EBS method can in fact be used for any coupling dependency in the system coordinate. The only constraint on system-bath coupling terms is that they need to be polynomial in the bath coordinates. Hence, any interaction of the following form may be considered by the EBS method:

\begin{equation}
    \hat{H}_{\rm SB} = \sum_{k,j}\sum_{l_k, l_j} \hat{f}_{k,j,l_k,l_j}(\hat{Q}_1) \times \hat{Q}_k^{l_k}\hat{Q}_j^{l_j}\, .
\end{equation}
Note that accessing nonlinear coupling terms is critical for molecular applications as quadratic bath terms ($\hat{Q}_k^2$ and $\hat{Q}_j\hat{Q}_k$) are essential to describe internal energy transfers, such as intramolecular vibrational redistribution (IVR).\cite{stannard_intramolecular_1981,bondybey_relaxation_1984, Nesbitt_IVR_1996}

\subsection{Effective bath states}
\begin{figure}[t]
	\centering\includegraphics[width=1\linewidth]{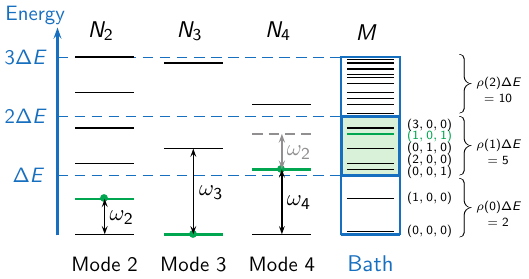}
	\caption{Representation of the coarse-graining procedure of the bath for a model system containing four modes (mode 1 is the system). \textit{Left:} Energy scale where the first three energy grains are indicated. \textit{Center:} energy levels of the three harmonic bath modes with their respective frequencies $\omega_k$ and their maximum number of quanta $N_k$. \textit{Right:} Transformation of the bath modes into a single ladder of effective energy states $\ket{m}$ (in blue). Individual microstates are represented in black inside the EES that contains them, with a label indicating the corresponding set of quantum numbers $\left(n_2,n_3,n_4\right)$. The number $\rho(m)\Delta E$ of microstates in each EES is also indicated. One specific bath state with one quantum of energy in mode 2 and one in mode 4 is highlighted in green. The position of the associated microstate $\left(1,0,1\right)$ is represented in green inside the effective state ladder. As shown in grey, its position is obtained by summing the energies of the individual bath modes. The coarse-graining procedure allows the basis set size to drop from $N_2\times N_3\times N_4$ microstates to a much smaller number $M$ of effective states. }
\label{fig:method_bath_graining} 
\end{figure}

The EBS method relies on the use of global effective energy states representing the total bath energy. This allows for a strong reduction of the bath dimension, a necessary step to deal with large baths. 
As detailed in Ref.~\onlinecite{Attal_EBS_model_2024}, the bath energy is discretized using an energy grain $\Delta E$, and a given effective bath state $\ket{m}$ is defined as the global bath state containing all microstates $\ket{\bf n} = \ket{n_2, \hdots, n_g}$ such that

\begin{equation}
m\Delta E 	\;\leq \;E({\bf n}) \;< \;(m+1)\Delta E\, .
\label{eq:def_bin}
\end{equation}
We denote by ${\bf n}\in m$ the fact that $\ket{\bf n}$ satisfies Eq.~\eqref{eq:def_bin}. 
The energy of a microstate ${\bf n}\in m$ is rounded down to $m\Delta E$. Since an effective state $\ket{m}$ contains several microstates, we introduce the bath density of state (DOS) $\rho(m)$ such that $\ket{m}$ contains $\rho(m) \Delta E$ microstates. 
The transformation from a bath made of $g-1$ harmonic oscillators to a coarse-grained ladder of EES with energy $m\Delta E$ and DOS $\rho(m)$ is illustrated in Fig.~\ref{fig:method_bath_graining}, on an example where $g=4$. 

The coarse-graining procedure assumes fast energy redistribution within each energy grain of the bath, and all microstates $\ket{\bf n}$ within a given state $\ket{m}$ are thus considered to be equiprobable.
As detailed in  Ref.~\onlinecite{Attal_EBS_model_2024} a microstate $\ket{\bf n}\in \ket{m} $ is hence replaced by
\begin{equation}
    \frac{1}{\sqrt{\rho(m)  \Delta E}} \ket{m}\,.
\end{equation}
This leads to the following effective bath Hamiltonian\cite{Attal_EBS_model_2024}

\begin{align}
\hat{H}_{\rm{B}} = \sum_{m = 0}^{M-1} m\Delta E \, \ket{m}\bra{m}\,,
\end{align}
where the number of effective bath states $M$ is directly related to the maximum bath energy:  $E_{\rm max} = M\Delta E$. Hence, the size of the effective basis set is determined both by the considered energy range and by the coarse-graining precision $\Delta E$, but it is independent of the number of bath modes $g-1$. 
The total number of states needed to describe the bath hence decreases from $\sim N^{g-1}$ microstates (if each bath mode is represented by $N$ states) to $M$ effective states, with $M \ll N^{g-1}$ and where $M$ does not depend on the number of modes (and hence on the molecule size).

Note that for numerical reasons,\cite{Attal_EBS_model_2024} the bath frequencies $\omega_k$ need to be rounded to the nearest multiple of $\Delta E$. We thus introduce the integer $m_k$ such that

\begin{equation}
    \hbar\omega_k = m_k \Delta E\, .
    \label{eq:rounded_freq}
\end{equation}
This procedure ensures that resonant processes are correctly described by the method and that they are not affected by numerical rounding errors. When $\Delta E$ is small enough, it does not significantly affect the frequency values. 

Note that all DOSs ($\rho$, $\rho^{(k)}$ and $\rho^{(j,k)}$) are computed using the Beyer-Swinehart algorithm,\cite{Beyer_Swinehart_1973, Stein_Rab_1973} which provides quantum DOSs for uncoupled harmonic oscillators. Owing to the rounding of the bath frequencies, this counting method is exact in our case, as long as $\Delta E$ is used as the grain size of the algorithm.Partial DOSs $\rho^{(k)}$  and $\rho^{(j,k)}$ are obtained by excluding mode $k$ (resp. modes $j,k$) from the counting when applying the Beyer-Swinehart algorithm.


\subsection{Coupling terms in the effective basis set}
The main technical difficulty in the EBS formalism resides in the calculation of the system-bath coupling Hamiltonian in the effective basis set $\ket{v,m}$. In this section, we detail the calculation of the coupling terms of Eq.~ q.~(\ref{eq:H_coupling_total}) in this basis set.  Since we have access to the system eigenstates, there is no difficulty in computing this part of the coupling. We thus focus on the bath matrix elements $\bra{m'}\hat{Q}_k^{l_k}\hat{Q}_j^{l_j}\ket{m}$. 

The coupling Hamiltonian of Eq.~(\ref{eq:H_coupling_total}) contains both two-mode and three-mode coupling terms. We first consider two-mode terms as they behave similarly to the linear case described in Ref.~\onlinecite{Attal_EBS_model_2024}. The additional difficulties arising for three-mode terms will then be detailed. 
 
\subsubsection{Two-mode coupling terms}
In the quartic expansion of Eq.~(\ref{eq:H_coupling_total}), two-mode coupling terms are either linear ($\hat{Q}_1^2\hat{Q}_k$) or quadratic ($\hat{Q}_1\hat{Q}_k^2$ and $\hat{Q}_1^2\hat{Q}_k^2$) in the bath coordinates. In both cases, only one bath mode $k$ is involved in the coupling. The main difficulty is to take into account coupling terms that act only on one bath mode using global effective bath states that are delocalized over the whole bath.
\\

\paragraph{Partition of the bath energy }
To do so, the bath energy $m\Delta E$ is divided in two parts:

\begin{equation}
m\Delta E = n_k \hbar \omega_k + E^*_{n_k}\,,
\label{eq:bin_m}
\end{equation}
with $n_k \hbar \omega_k$ the energy of mode $k$ when there are $n_k$ quanta in this mode, and $E^*_{n_k}$ is the energy accessible to the other bath modes. This energy will be referred to as the spectator energy, since it corresponds to the energy shared by the spectator bath modes $j \ne k$.
By definition, the spectator energy $E^*_{n_k}$ is not affected by the transitions induced by $\hat{Q}_k$. An effective state $m^*$ can be associated with $E^*_{n_k}$, and the expression ${\bf n}^* \in m^*$ will be used to indicate that the spectator microstate ${\bf n}^* = \{ n_j\}_{j\ne k}$ has an energy between $m^*\Delta E$ and  $(m^*+1)\Delta E$.
For a given value of $m$ and $n_k$, there are multiple possibilities to distribute $E^*_{n_k}$ between the spectator bath modes. A spectator mode DOS $\rho^{(k)}(m^*)$ is thus introduced. It counts the number of spectator microstates ${\bf n}^*$ in state $\ket{m^*}$. 
\\

\paragraph{Linear terms in the bath coordinates}
The case of linear coupling terms in the bath coordinates has been treated in detail in a previous article.\cite{Attal_EBS_model_2024} Here, we only summarize the main steps of the calculation. 

Operator $\hat{Q}_k$ only affects mode $k$ for which it induces transitions between $n_k$ and $n'_k=n_k\pm1$. Upon applying $\hat{Q}_k$  to an effective state $\ket{m}$, the bath energy becomes

\begin{equation}
m'\Delta E = n'_k \hbar \omega_k + E^*_{n_k}\,.
\label{eq:bin_m_prime}
\end{equation}
Such transitions are characterized by a parameter $\Delta m_k~=~|m' - m|$ that only depends on $\omega_k$ and $\Delta n_k~=~ n'_k - n_k$, which is equal to $\pm 1$ for a linear coupling. Operator $\hat{Q}_k$ couples a state $\ket{m}$ to the states $\ket{m+\Delta m_k}$ and $\ket{m-\Delta m_k}$ such that 

\begin{equation}
\Delta m_k = \left|\frac{\Delta n_k \hbar \omega_k}{\Delta E}\right| = |\Delta n_k \times m_k| \;.
\label{eq:def_delta_m}
\end{equation}
Note that for a linear coupling $\Delta m_k = m_k$. 
The total coupling between $\ket{m}$ and $\ket{m\pm\Delta m_k}$ is obtained by adding the contributions of all the states $n_k$ accessible to mode $k$ when the bath has an energy $m\Delta E$, meaning all the $n_k$ such that $n_k\hbar \omega_k \leq m\Delta E$. The maximum integer fulfilling this condition for a given $m$ is defined as
\begin{equation}
	N_k(m)= \left[\frac{m\Delta E}{\hbar \omega_k} \right] = \left[\frac{m}{m_k}\right],
	\label{eq:N_k}
\end{equation}
where $[x]$ denotes the rounding to the nearest integer from $x$. The total coupling is obtained by summing over all the $n_k$ such that $0\leq n_k \leq N_k(m)$. 

The expression of operator $\hat{Q}_k$ in the effective bath basis set $\ket{m}$ is obtained by\cite{Attal_EBS_model_2024}
\begin{enumerate}
\item Re-writing the microstates $\ket{\bf n}$ as $\ket{n_k,{\bf n}^*}$ to separate mode $k$ from the spectator bath modes.
 \item Computing the matrix elements of $\hat{Q}_k$ in the harmonic basis set:

 \begin{equation}
\bra{{\bf n'}} \hat{Q}_k \ket{{\bf n}} = \bra{n'_k} \hat{Q}_k \ket{n_k}\times\delta_{n'_k, n_k \pm 1}\times\delta_{{\bf n'}^*,{\bf n}^*} \,.
 \end{equation}

\item Using a reordering of the sums adapted to the bath energy ladder:
    \begin{equation}
\sum_{{\bf n}} \ket{\bf n} = \sum_{m = 0}^{M-1} \sum_{{\bf n}\in m} \ket{\bf n}= \sum_{m = 0}^{M-1} \sum_{n_k = 0}^{N_k(m)} \sum_{{\bf n}^* \in m^*}\ket{n_k,{\bf n}^*}\, .
\end{equation}

\item Counting the number of accessible microstates in each EES $\ket{m}$ to obtain the microcanonical probability $\mathbb{P}(m,n_k)$ of having $n_k$ quanta in mode $k$ while being in state $\ket{m}$. This microcanonical probability is given by\cite{Attal_EBS_model_2024}

\begin{equation}
     \mathbb{P}(m,n_k) = \frac{\rho^{(k)}(m-n_k m_k)}{\rho(m)}\, ,
\end{equation}
This expression holds under the assumption that the energy redistribution inside each EES $\ket{m}$ is much faster than the typical system-bath interaction time.
\end{enumerate}
The above procedure leads to the effective operator\cite{Attal_EBS_model_2024}

\begin{multline}
\hat{Q}_{k}  = \sum_{m = 0}^{M-\Delta m_k-1}  \sum_{n_k = 0}^{N_k(m)} \sqrt{\frac{\hbar (n_k+1)}{2\omega_k}} \; \mathbb{P}(m,n_k)\\ \times \left(\ket{m+\Delta m_k}\bra{m} + \ket{m}\bra{m+\Delta m_k}\right)\,,
\label{eq:transition_plus}
\end{multline}
where the condition $m \leq M-\Delta m_k-1$ ensures that $m+\Delta m_k$ remains below the highest considered effective state ($M-1$).
\\

\paragraph{Quadratic terms in the bath coordinates}

Quadratic terms $\hat{Q}_1\hat{Q}_k^2$ and $\hat{Q}_1^2\hat{Q}_k^2$ are transcribed in the effective basis set in the same way as linear terms, except that $\hat{Q}_k^2$ induces transitions with $\Delta n_k = \pm 2$. Hence,  $\Delta m_{k}= 2\,m_k$ (see Eq.~\eqref{eq:def_delta_m}). 
The quadratic operator also generates terms that conserve the population in a given harmonic state $n_k$ (i.e. $\Delta n_k =0$).
This part of the operator can be written as
\begin{equation}
\sum_{m = 0}^{M-1}  \sum_{n_k = 0}^{N_k(m)}\bra{n_k} \hat{Q}_k^2\ket{n_k}\mathbb{P}(m,n_k) \ket{m}\bra{m}.
\end{equation}
It has a diagonal form since it does not induce any transition inside the bath.
The effective expression of operator $\hat{Q}_k^2$  in the EES basis set is hence given by
    \begin{equation}
    \begin{aligned}  
\hat{Q}^{2}_{k} =& \sum_{m = 0}^{M-1}  \sum_{n_k = 0}^{N_k(m)}\frac{\hbar}{2\omega} (2n_k + 1)\,\mathbb{P}(m,n_k) \ket{m}\bra{m}
\\ &+\sum_{m = 0}^{M-\Delta m_k-1}  \sum_{n_k = 0}^{N_k(m)}\frac{\hbar}{2\omega} \sqrt{(n_k+2)(n_k+1)}\,\mathbb{P}(m,n_k) \\ 
&\quad\quad \quad \times \left(\ket{m+\Delta m_k}\bra{m} + \ket{m}\bra{m+\Delta m_k}\right)\, .
\end{aligned}
\end{equation}

\subsubsection{Three-mode coupling terms}
Three-mode coupling terms $\hat{Q}_1\hat{Q}_j\hat{Q}_k$ are more complex to deal with since they involve two different bath modes $j$ and $k$. This leads to several difficulties.
First, the bath energy needs to be split into three parts: the energy in mode $j$, the energy in mode $k$, and the remaining spectator energy that is distributed between bath modes $l  \notin \{j,k\}$. This notably implies that additional DOSs need to be computed.
Three-mode terms can induce two types of transitions: either both modes $j$ and $k$ gain (or lose) a quantum of energy, or one mode gains a quantum of energy while the other loses one. 
In the latter case, the enumeration of the accessible microstates in each EES requires particular care, and the effective states involved in the coupling terms are not always straightforward.
\\

\paragraph{Partition of the bath energy}
The bath is assumed to have an energy $m\Delta E$, and both modes $j$ and $k$ are isolated from the other bath modes by writing

\begin{equation}
m\Delta E = n_j \hbar \omega_j+ n_k \hbar \omega_k + E^*_{n_j, n_k}\, ,
\label{eq:bin_m_3modes}
\end{equation}
where the spectator energy $E^*_{n_j, n_k}$ is given by

\begin{equation}  
E^*_{n_j, n_k} = \sum_{l  \notin \{j,k\}} n_l\hbar \omega_l = m^*_{j,k} \Delta E\,.
\label{eq:Spectator_nrj_3modes}
\end{equation}
This energy is associated with a EES $m^*_{j,k}$ such that

\begin{equation}
    m^*_{j,k} = m - n_jm_j - n_km_k\,.
\end{equation}
As in the two-mode case, there are many possibilities to distribute the spectator energy between the spectator modes $l  \notin \{j,k\}$. The possible distributions of this energy are obtained by excluding modes $j$ and $k$ that have a fixed energy $n_j \hbar \omega_j$ and $n_k \hbar \omega_k$, respectively. This leads to a new DOS $\rho^{(j,k)}(m^*_{j,k})$,  which counts the number of spectator microstates $\{n_l\}_{l \notin \{j,k\}}$ that have an energy $m^*_{j,k}\Delta E$. 

\begin{figure*}[t]
    \centering
    \includegraphics[width=.9\linewidth]{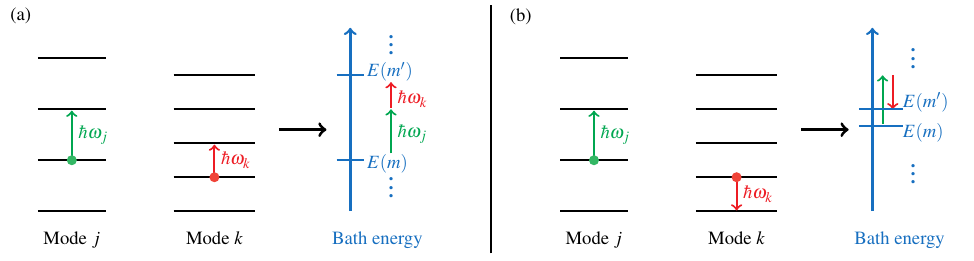}
     \caption{Illustration of possible transitions involving two bath modes $j$ and $k$, with frequencies $\omega_j > \omega_k$. \textit{Left:} a combination band where modes $j$ and $k$ both gain one quantum of energy, leading to a transition energy $\hbar\omega_j+\hbar\omega_k$. \textit{Right:} a difference band where mode $j$ gains one quantum of energy while mode $k$ loses one, leading to a transition energy $\hbar\omega_j-\hbar\omega_k$.}
    \label{fig:difference_bands}
\end{figure*}

When the bath has an energy $m\Delta E$, mode $j$ has access to all energy levels $n_j$ such that $n_j \hbar\omega_j \leq m\Delta E$, hence to all $n_j$ such that $0 \leq n_j \leq N_j(m)$. If there are $n_j$ quanta in mode $j$, then mode $k$ only has access to the energy levels such that $n_k \hbar\omega_k \leq m\Delta E - n_j \hbar\omega_j$. Meaning that the maximum value of $n_k$ is

\begin{equation}
N_k(m -n_jm_j)= \left[\frac{m\Delta E-n_j \hbar \omega_j}{\hbar \omega_k} \right] = \left[\frac{m - n_jm_j}{m_k}\right]\,.
\end{equation}
This integer will be denoted as $N_k(m,n_j)$. Similarly, we define the probability $\mathbb{P}(m,n_j,n_k)$ of having $n_j$ quanta in $j$ and $n_k$ quanta in $k$ while being in state $\ket{m}$. This microcanonical probability is given by

\begin{equation}
    \mathbb{P}(m,n_j,n_k) = \frac{\rho^{(j,k)}(m - n_jm_j - n_km_k)}{\rho(m)}\; . 
\end{equation}

Since two bath modes are involved in the transition, they can either both gain (or lose) one quantum of energy ($n_j\to~n_j \pm1$ and $n_k\to n_k \pm1$) or one can gain one quantum of energy while the other loses one ($n_j\to n_j \pm1$ and $n_k\to n_k \mp 1$). The operator $\hat{Q}_j\hat{Q}_k$ can hence be decomposed in four contributions:

\begin{equation}
    \hat{Q}_j\hat{Q}_k =\hat{Q}_{j}^{+}\hat{Q}_{k}^{+} + \hat{Q}_{j}^{-}\hat{Q}_{k}^{-} + \hat{Q}_{j}^{+}\hat{Q}_{k}^{-} + \hat{Q}_{j}^{-}\hat{Q}_{k}^{+}\; ,
\end{equation}
where the exponent $\pm$ indicates whether the associated mode gains or loses one quantum of energy.
The spectral signatures of operators $\hat{Q}_{j}^{+}\hat{Q}_{k}^{+}$ and $\hat{Q}_{j}^{-}\hat{Q}_{k}^{-}$ are called combination bands, and operators $\hat{Q}_{j}^{+}\hat{Q}_{k}^{-}$ and $\hat{Q}_{j}^{-}\hat{Q}_{k}^{+}$ give rise to difference bands. In the rest of this section, we will use this terminology and rely on an analogy with infrared spectroscopy to explain the calculation of three-mode coupling terms. These coupling operators do not necessarily describe radiative processes -- they can also account for, e.g., internal energy transfers -- and the aim of the analogy with vibrational spectroscopy is to help the reader interpret and visualize their action on the effective bath states.
\\

\paragraph{Combination bands}
Combination bands behave much like the two-mode transitions described earlier, with a transition parameter

\begin{equation}
    \Delta m^+_{jk}= \left[\frac{\hbar (\omega_j+\omega_k)}{\Delta E} \right] = m_j + m_k.
\end{equation}
As illustrated in Fig.~\ref{fig:difference_bands}(a), if $n_j\to n_j +1$ and $n_k\to n_k +1$, then the bath gains an energy $\hbar\omega_j+\hbar\omega_k$ and the transition frequency is the sum of the frequencies of both modes.
A similar procedure to the one of Ref.~\onlinecite{Attal_EBS_model_2024} then leads to the following effective coupling operator

\begin{multline}
\hat{Q}_{j}^{+}\hat{Q}_{k}^{+} = \sum_{m = 0}^{M -\Delta m^+_{jk}-1}  \sum_{n_j = 0}^{N_j(m)}\sum_{n_k = 0}^{N_k(m,n_j)}\frac{\hbar\sqrt{(n_j+1)(n_k+1)}}{2\sqrt{\omega_j\omega_k}}
\\ \times\mathbb{P}(m,n_j,n_k) \ket{m+\Delta m^+_{jk}}\bra{m}.
\label{eq:transition_plus_3modes}
\end{multline}
Operator $\hat{Q}_{j}^{-}\hat{Q}_{k}^{-}$ can be obtained as the Hermitian conjugate of $\hat{Q}_{j}^{+}\hat{Q}_{k}^{+}$.
\\

\paragraph{Difference bands}
A difference band occurs when one of the two involved modes gains energy while the other one loses some. Such a transition is governed by the following parameter

\begin{equation}
    \Delta m^-_{jk}= \left[\frac{\hbar (\omega_j-\omega_k)}{\Delta E} \right] = m_j - m_k\, ,
    \label{eq:def_delta_m_minus}
\end{equation}
which can be positive or negative depending on the sign of $\omega_j-\omega_k$. A transition $m\to \ket{m+\Delta m^-_{jk}}$ corresponds to $(n_j,n_k) \to (n_j+1,n_k-1)$, and a transition $m\to \ket{m-\Delta m^-_{jk}}$ corresponds to $(n_j,n_k) \to (n_j-1,n_k+1)$. However, as the sign of $\Delta m^-_{jk}$ is unknown, the transition towards $\ket{m+\Delta m^-_{jk}}$ (resp. $\ket{m-\Delta m^-_{jk}}\,$) does not necessarily represent a total energy gain (resp. loss) for the bath. 

In the following we consider a transition $(n_j,n_k) \to (n_j+1,n_k-1)$ such as the one shown in Fig.~\ref{fig:difference_bands}(b). 
For such a transition to be allowed, there must initially be at least one quantum of energy in mode $k$; this leads to two constraints: $n_k \geq 1$ and $m \geq m_k = \hbar\omega_k/\Delta E$. It also implies that he maximum quantum level $N_j$ reached by mode $j$ in a given state $m$ is such that $ N_j\hbar\omega_j + \hbar\omega_k = m\Delta E$. Leading to

\begin{equation}
\begin{aligned}
    N_j\hbar\omega_j &= m\Delta E -\hbar\omega_k   \\
    &=m\Delta E -m_k \Delta E  \\
    &=(m - m_k)\Delta E.
\end{aligned} 
\end{equation}
The maximum value reached by $n_j$ in $m$ is, hence, not given by  $N_j(m)$ but by $N_j(m-m_k)$. Then, the maximum value of $n_k$ is given by $N_k(m,n_j) = N_k(m-n_jm_j)$. Moreover, since the effective state containing $n_j +1$ must remain below the highest bath state $M-1$, the initial state $m$ must be smaller than $M-m_j-1$. In the end, the various sums involved in the effective expression of $\hat{Q}_{j}^{+}\hat{Q}_{k}^{-}$ are constrained by

\begin{align}
    \left\{\begin{array}{lll}
    0\leq n_j\leq N_j(m-m_k)  \\
    1\leq n_k\leq N_k(m,n_j) \\
    m_k \leq m \leq M - m_j - 1\,, \end{array}\right.
\end{align}
Simple but fastidious calculations based on these ideas (see App.~\ref{AppA}) lead to the following effective expression for the difference bands operator
\begin{multline}
\hat{Q}_{j}^{+}\hat{Q}_{k}^{-} = \sum_{m = m_k}^{M -m_j-1}  \sum_{n_j = 0}^{N_j(m-m_k)}\sum_{n_k = 1}^{N_k(m,n_j)}\frac{\hbar\sqrt{(n_j+1)n_k}}{2\sqrt{\omega_j\omega_k}}
\\ \times \mathbb{P}(m-m_k,n_j,n_k-1) \ket{m+\Delta m^-_{jk}}\bra{m}\,. 
\end{multline}
Operator $\hat{Q}_{j}^{-}\hat{Q}_{k}^{+}$ is obtained as the Hermitian conjugate of the above operator.

\subsection{Intramolecular vibrational redistribution}
Once the total effective Hamiltonian $\hat{H} = \hat{H}_{\rm S}+\hat{H}_{\rm B}+\hat{H}_{\rm SB} $ is obtained, the time-dependent Schrödinger equation (TDSE) can be solved starting from an initial state $\ket{v_0,m_0}$, \textit{i.e.} the system starts in a given vibrational state $v_0$ and the bath is at an energy $m_0\Delta E$. By propagating the total effective wavefunction and computing the relevant observables, the redistribution of the energy and populations after excitation of the mode of interest (i.e. the system) can be followed through time. 

The effective Hamiltonian being time-independent, the TDSE can be solved exactly and the wavefunction $\ket{\psi(t)}$ can be obtained at any time $t$ as a function of the eigenenergies of $\hat{H}$ and of the coefficients of the initial state. 
The populations of the system and bath states ($\ket{v}$ and $\ket{m}$, respectively) can then be computed along a given trajectory:

\begin{align}
{P}_v(t) &= \sum_m |\braket{v,m|\psi(t)}|^2 
\label{eq:probas_t_v}\,,\\ 
 {P}_m(t) &= \sum_v |\braket{v,m|\psi(t)}|^2 \,.
 \label{eq:probas_t_m}
\end{align}
In a molecular context, these observables enable us to follow IVR after the initial excitation of a mode of interest.

The bath can be prepared at 0~K by starting from its ground state $\ket{m_0 = 0}$, but it can also be initialized at a non-zero temperature. To do so, many energy-fixed trajectories starting from different initial bath energies $E_0$ (i.e. from different EESs $\ket{m_0}$) are performed. Then, they are re-weighted by the corresponding thermal probability to recover the canonical observable $A(T)$ from the microcanonical observable $A(E)$:

\begin{align}
\label{eq:thermal_p}
A(T) &= \frac{1}{Z(T)}\int A(E)\rho(E)e^{-E/k_{\rm B}T} dE, \\
Z(T) &= \int \rho(E)e^{-E/k_{\rm B}T}dE,\nonumber 
\end{align}
where $k_{\rm B}$ and $Z(T)$ denote the Boltzmann constant and the canonical partition function, and where $\rho(E)$ corresponds to the DOS at energy $E$. 
Note that this thermal averaging \textit{only applies to the initial state of the bath.} The system is still prepared in a given eigenstate $\ket{v}$, and the dynamics remains microcanonical since the bath is not in contact with the thermostat for times $t > 0$.

\subsection{Temperature resolved infrared absorption spectroscopy}
In the EBS method, only the mode of interest is coupled to the external field ${\bf E}_0$ and a linear approximation is used for the dipole moment \mbox{\boldmath$\mu$}. Assuming a large band excitation, the transition probability per time unit between two eigenstates $\ket{\varphi_\alpha}$ and $\ket{\varphi_\gamma}$ of the effective Hamiltonian is given by

\begin{equation}
    \begin{aligned}
I_{\alpha,\gamma} (\omega_{\alpha \gamma}) &\propto\ \mid \bra{\varphi_\gamma}{\bf E}_0. \frac{\partial \mbox{\boldmath $\mu$}}{\partial Q_{1}}  \hat{Q}_{1} \ket{\varphi_\alpha}\mid^2 \\
&\propto \ \mid \langle \varphi_\gamma \mid \hat{Q}_{1}  \mid \varphi_\alpha\rangle \mid^2 \,,
\label{eq:spectrum}
\end{aligned}
\end{equation}
where $\hbar\omega_{\alpha \gamma} = E_\gamma - E_\alpha$ is the energy difference between the two states.  Because of the linear form of the dipole assumed here, only transitions with $\Delta v= \pm 1$ are correctly reproduced. Overtone transitions ($|\Delta v| > 1$) are only weakly allowed by the anharmonicity of the system, and higher orders of the transition dipole would be required to obtain correct intensities for them. 

The temperature-resolved infrared spectrum, taking into account both the absorption and stimulated emission, is given by

\begin{multline}
I^{(\beta)} (\omega) =  \frac{\hbar\omega}{Z(\beta)} \sum_{\alpha} \sum_{\gamma >\alpha} (e^{-\beta E_\alpha}-e^{-\beta E_\gamma})  \\ \times I_{\alpha,\gamma}(\omega_{\alpha \gamma}) \, \times  \delta (\omega -\omega_{\alpha \gamma })\,,
\end{multline}
with $\beta  = 1/k_B T$ and  $Z(\beta)= \sum_{\alpha} e^{-\beta E_\alpha}$.  
The spectrum is convoluted with a Gaussian function of chosen full width at half maximum (FWHM), which mimics the broadening of the transitions due to the molecule rotation and the laser linewidth. 

Once a spectrum is obtained, it is possible to identify transitions from the involved eigenstates  $\ket{\varphi_\alpha}$ and $\ket{\varphi_\gamma}$. To do so, the eigenstates are decomposed in the $\ket{v,m}$ basis set where physical interpretation is more natural. The largest coefficients in their decomposition are used to interpret and assign the transition. 
Note that for the bath, only the effective state $\ket{m}$ is known, not the specific microstate involved in the transition. Once the value of $m$ is known, we look at the possible bath microstates that have an energy $m\Delta E$. This leads to a restricted list of possible microstates involved in the transition. Complementary information, such as the symmetries of the modes, the value of the coupling terms or the possible resonances, can then help to narrow down the possibilities and assign the transition.

\section{Benchmark on a 10-mode model system}
\label{sec:Spectra_10modes}

To test the ability of the EBS method to compute absorption spectra, we first perform calculations on a 10-mode model system made of frequencies and coupling parameters that we generated to have the same orders of magnitude as a realistic vibrational system, albeit with relatively strong anharmonicities to create a richer spectrum.

\subsection{Computational details}

\begin{table}[!b]
    \centering
    \begin{tabular*}{1\linewidth}{@{\extracolsep{\fill}} c|cccccccccc}
    \hline \hline
     \rule[.1ex]{0pt}{2.5ex}$i$ &1 &2&\textcolor{ForestGreen}{\textbf{3}}&4&5&6&7&8&9& 10 \\ [0.1cm] \hline
    \rule[.1ex]{0pt}{2.5ex}$\;\omega_i\;$& 410& 560&\textcolor{ForestGreen}{\textbf{800}}&830&1260&1450&1510&1660&1712&1860\\ [0.1cm]
         \hline \hline
    \end{tabular*}
    \caption[Harmonic frequencies of the 10-mode model system]{Harmonic frequencies of the 10-mode model system, given in cm$^{-1}$. The mode of interest is emphasized in green.}
    \label{tab:freq_10_modes}
\end{table}

The 10-mode model system is represented by a quartic PES such as the one of Eq.~\eqref{eq:quartic_PES}. 
The harmonic frequencies of the 10 modes are given in Table~\ref{tab:freq_10_modes} and the anharmonic coefficients $\alpha$ and $\beta$ used in $\hat{H}_{\rm S}$ and $\hat{H}_{\rm SB}$ are provided in App.~\ref{App6}.  As emphasized in Table~\ref{tab:freq_10_modes}, the chosen mode of interest for the calculations presented below is $i_0 =3$, which has a harmonic frequency of $\omega_3 = 800$~cm$^{-1}$. Note that all the coupling terms involving only bath modes (i.e., only modes $k\ne i_0$) are neglected in the calculation since the EBS method assumes the bath to be made of uncoupled harmonic oscillators. However, couplings connecting the mode of interest $i_0 =3$ to individual bath modes $k\ne i_0$ are considered in the calculations. The system's eigenenergies and eigenstates are computed using variational basis representation.\cite{Gatti2009,Loise_these_2024} The fundamental frequency of $i_0$ is found at $\omega_{0\to1} = 794$~cm$^{-1}$, and its two first hot bands at $\omega_{1\to2} = 788$~cm$^{-1}$ and $\omega_{2\to3} =782$~cm$^{-1}$, respectively. 
As in Eq.~\eqref{eq:spectrum}, the dipole moment is assumed to be linear, i.e. $\mu(Q_1) = a_0 + a_1Q_1$. No permanent dipole was considered ($a_0 = 0$), and since intensities are computed up to a multiplicative factor, the prefactor $a_1$ is not relevant to the calculation (see Eq.~\eqref{eq:spectrum}).

The rather small size of this test system was chosen in order to compare the EBS results with full-dimensional calculations. These calculations were performed using the exact same system and parameters as for the EBS calculations. In particular, the bath modes are also assumed to be harmonic and uncoupled in the full-dimensional calculation. To reduce the basis set size, an energy criterion was used in the full-dimensional calculation, and only the bath microstates with an energy under a certain threshold $E_c$ were kept in the calculations. 

The EBS parameters used for the calculations discussed bellow are $N_v = 5$, $M = 3\,500$, and $\Delta E = 1$~cm$^{-1}$. Hence the maximal bath energy is $M\Delta E = 3\,500$~cm$^{-1}$. However, for such a small bath with only 9 modes, many bath energy grains are empty in the sense that there is no bath microstate such that Eq.~\eqref{eq:def_bin} is fulfilled. The corresponding EES are unphysical and hence excluded from the calculations.\cite{Attal_EBS_model_2024} In practice we only construct a reduced effective Hamiltonian with $M_{\rm eff} < M$ bath states. In the present case, $M_{\rm eff} =143$. 
For the full-dimensional calculation, $N_v = 5$ states were considered for the system and 225 bath states with an energy below $E_c = 3\,500$~cm$^{-1}$ were used. In both cases, the spectra were convoluted with a Gaussian function having an FWHM of 2.5~cm$^{-1}$.

\subsection{Infrared absorption spectra}
\label{subsec:spectra_10modes}

\begin{figure}[t]
    \centering
    \includegraphics[width=1\linewidth]{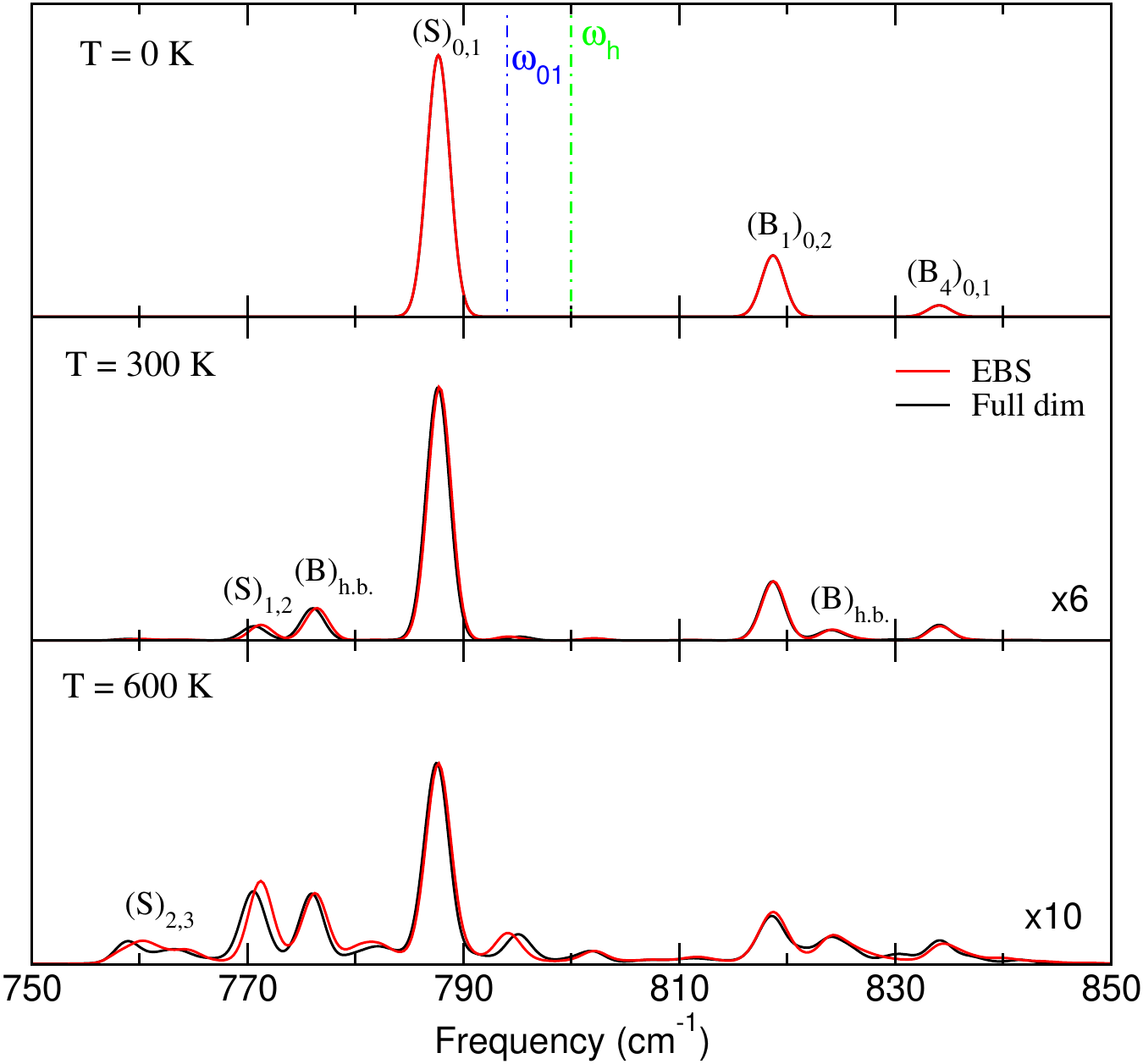}
    \caption{Absorption spectra of the 10-mode model system in the spectral region of the mode of interest, obtained at 0, 300 and 600~K. The EBS and full-dimensional results are superimposed in red and black, respectively. Both spectra have been re-normalized to have the same maximal intensity. The system transitions $v \to v'$ are denoted as (S)$_{v,v'}$. The transitions $n_k \to n'_k$ of a given bath mode $k$ are denoted as (B$_k$)$_{n_k,n_k'}$. When there are several bath hot bands involved in a feature, it is denoted as (B)$_{\rm h.b.}$. In the upper panel, the mode of interest harmonic ($\omega_h$) and fundamental ($\omega_{01}$) frequencies are indicated by vertical dashed lines.
    }
    \label{fig:Spectra_10_modes}
\end{figure}

The infrared absorption spectra at different temperatures obtained using the parameters above are shown in Fig.~\ref{fig:Spectra_10_modes}, where EBS results are compared with the corresponding full-dimensional calculations. Both methods provide exactly the same 0~K frequencies, and the agreement is almost perfect at 300~K. Even at 600~K, both methods agree very well, and even though small differences can be seen, the maximal shift between the two spectra is of $\sim 1$~cm$^{-1}$. This value corresponds to the bath energy grain $\Delta E$ used in the EBS calculations and is, hence, the best accuracy we could expect with the current parameters. There are also small intensity differences in the shifted bands. 
\\

\paragraph{Analysis of the 0~K spectrum}

As seen in the top panel of Fig.~\ref{fig:Spectra_10_modes}, even at 0~K, the mode of interest undergoes three different transitions. This is due to the anharmonic couplings between the mode of interest and various bath modes. As expected at 0~K, the three transitions start from the absolute ground state of the molecule ($\ket{\varphi_0} = \ket{v=0, m = 0}$). The most intense 0~K transition mostly corresponds to the fundamental transition of the mode of interest ($v = 0\to v = 1$). Its frequency is shifted to the red from both its harmonic frequency ($-12$~cm$^{-1}$) and its fundamental frequency ($-6$~cm$^{-1}$). The difference between the harmonic and fundamental frequencies of the mode of interest comes from its intra-mode anharmonicity. The further shift from its fundamental frequency obtained in Fig.~\ref{fig:Spectra_10_modes} comes from the coupling to other modes, thus showing that the EBS method is able to reproduce both intra-mode and inter-mode anharmonicities.  The final state of this transition is actually not $\ket{v = 1, m = 0}$ but a mixture of this state (76\%) with the bath excited states $\ket{v = 0,m = \hbar\omega_1/\Delta E}$ (20\%) --- denoted as $\ket{v = 0,m = \omega_1}$ in the following --- and $\ket{v = 0,m = \omega_4}$ (2\%). 

The two other transitions observed at 0~K are also a mix of system and bath transitions. Interestingly, in both cases, the bath excited state is predominant in the final state, even though the bath modes are not coupled to the field. These transitions are thus only possible if bath modes borrow intensity from the system. Their presence indicates that there are strong anharmonic interactions between the system and the bath. The agreement between full-dimensional calculations and EBS results shows that our method is able to correctly capture such strongly anharmonic behaviors. 

More precisely, the transition denoted as (B$_1$)$_{0,2}$ in Fig.~\ref{fig:Spectra_10_modes} corresponds to the first overtone of mode 1 ($n_1 = 0\to n_1 = 2$) and the one labeled (B$_4$)$_{0,1}$ to the fundamental transition of mode 4 ($n_4 = 0 \to n_4 = 1$). 
The predominance of bath modes 1 and 4 in the spectrum comes from the quasi-resonant condition with the fundamental transition of the mode of interest:

\begin{equation}
   \omega_{0\to 1} \approx 2\omega_1 \approx \omega_4  \,, 
   \label{eq:quasi_resonance}
\end{equation}
with $\omega_{01} = 794$~cm$^{-1}$, $\omega_1 = 410$~cm$^{-1}$ and $\omega_4 = 830$~cm$^{-1}$. The (quasi) Fermi resonance\cite{Fermi_1931} between the system and mode 1 induces a strong interaction between the fundamental transition of the mode of interest and the first two-quanta transition of mode 1. Such a transition is only accessible in calculations when including coupling terms that are quadratic in the bath coordinates, namely $Q_{i_0}Q_1^2$ for this specific transition. The analysis of the 0~K spectrum thus emphasizes the importance of including quadratic coupling terms in the bath coordinates to correctly reproduce the behavior of molecule-like systems.
\\

\paragraph{Temperature effects}
\begin{table}[!b]
    \centering
    \begin{tabular*}{1\linewidth}{@{\extracolsep{\fill}} c|cccccc}
    \hline \hline
    \rule[.1ex]{0pt}{2.8ex}$\ket{v,m}$ & $\ket{0,0}$ &  $\ket{0,\omega_1}$& $\ket{0,\omega_2}$ & $\ket{1,0}$ & $\ket{0,2\omega_1}$ &  $\ket{0,\omega_4}$  \\ [0.15cm] \hline
    \rule[.1ex]{0pt}{2.8ex}$E_{\rm EBS}$ & 0 & 410 & 560 & 794 & 820 & 830\\ [0.1cm]\hline
    \rule[.1ex]{0pt}{2.8ex}${\rm P}(300 \text{K})$ & 0.55 & 0.16 & 0.10 & 0.05 & 0.05 & 0.05 \\ [0.1cm]
         \hline \hline
    \end{tabular*}
    \caption{Most significantly populated states at 300~K, ordered by increasing EBS energy $E_{\rm EBS} = E_v + m\Delta E$ given in cm$^{-1}$. Boltzmann probabilities at 300~K are also given.}
    \label{tab:lowets_states_300K}
\end{table}

At 300~K new features appear, including the first hot band of the mode of interest ($v = 1 \to v = 2$) and several bath hot bands (see Fig.~\ref{fig:Spectra_10_modes}). Contrary to the 0~K case, several states are thermally populated at 300~K, the most populated ones are given in Table~\ref{tab:lowets_states_300K}. All except $\ket{0,\omega_2}$ contribute to the 300~K spectrum. This specific state does not appear in the spectrum because it does not induce any resonant interaction with the system in that spectral region. 

At 600~K, the second hot band of the system ($v = 2 \to v = 3$) starts forming, and more bath hot bands are visible. As could be expected, they are starting from higher initial states than the hot bands that were already visible at 300~K, with initial states like, e.g., $\ket{0, \omega_1 + \omega_4}$, $\ket{1, 2\omega_1}$, or $\ket{0, 4\omega_1}$.

\subsection{Intramolecular vibrational redistribution}

Internal population transfers after the initial excitation of the system, such as IVR, can also be obtained from the EBS method by following the time evolution of the system and bath populations. 
\\

\paragraph{Starting from the bath ground state}

\begin{figure}[t]
    \includegraphics[width=1\linewidth]{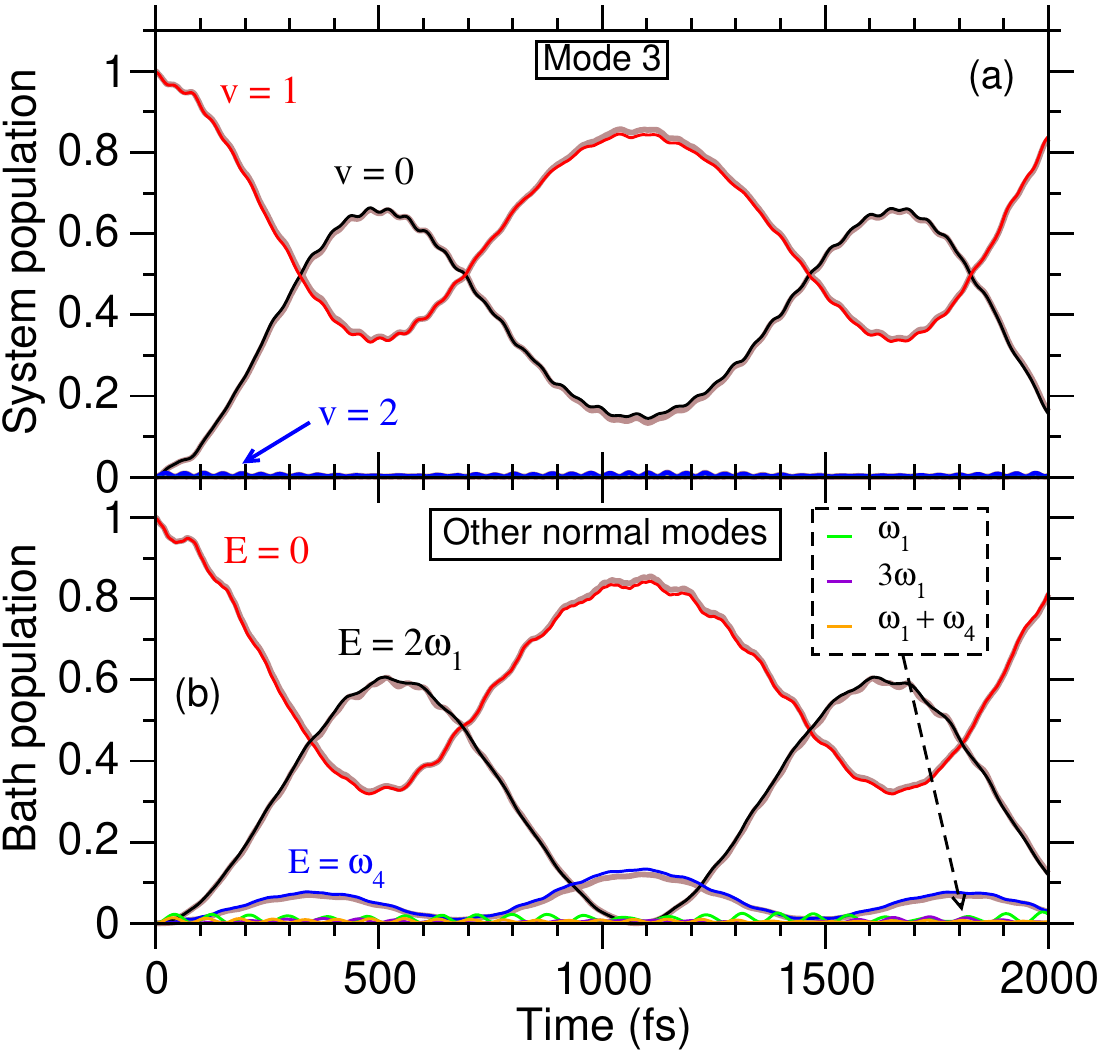}
	\caption{Time-evolution of the 10-mode model system starting from $\ket{v = 1, m = 0}$. (a) Evolution of the population in the vibrational states of the mode of interest $i_0 =3$. (b) Evolution of the population in the effective bath states, labeled by their energy. The bath contains all the modes except $i_0$. All the bath states gaining at least 1\% population are labeled. In both panels, the full-dimensional results are shown as thick brown lines.}
    \label{fig:IVR_10modes_v1_m0}
\end{figure}

The mode of interest  $i_0 = 3$ is prepared in its first vibrational excited state ($v = 1$), and the other modes of the molecule are assumed to initially be in their ground state. Hence, the bath starts in its ground state $m = 0$. The evolution of the system and bath populations along a 2~ps trajectory starting from this initial state $\ket{1,0}$ is shown in Fig.~\ref{fig:IVR_10modes_v1_m0}, where it is compared with full-dimensional calculations. The results from both methods coincide almost perfectly.

As seen in Fig.~\ref{fig:IVR_10modes_v1_m0}(a), the population in $v= 1$ first decreases as the system relaxes toward its ground state. However, only 65\% of the population is transferred to the ground state before a recurrence phenomenon starts occurring, with around 85\% of the population coming back in $v = 1$ after 1~ps. This recurrence phenomenon is typical of finite baths and can occur in real molecular systems.\cite{Morse_Surface_2012,Attal_EBS_model_2024}
The significant transfer of the system's energy to the bath is due to the quasi-resonance condition described in Eq.~\eqref{eq:quasi_resonance}.
Yet, this transfer is not complete because of the remaining detuning between the system and bath frequencies. A small population can also be seen in $v = 2$.  It comes from coupling terms beyond the rotating-wave approximation, which are included in the EBS method. These terms allow --- with a very small probability --- the system to simultaneously have an excitation from $v = 1$ to $v = 2$ and give a quantum of energy to the bath.\cite{fischer_hierarchical_2020, fischer_non-markovian_2022} 

Fig.~\ref{fig:IVR_10modes_v1_m0}(b) shows that the effective bath state with energy $E = 2\omega_1$, which is the closest to the resonance condition, is by far the one gaining the most population from the system's relaxation. The second closest effective state $E = \omega_4$ also gains a non-negligible population. As emphasized in the same panel, all the bath states gaining at least 1\% population during the trajectory are combinations of these two modes. This matches the conclusions made by analyzing the spectra. 
The analysis of the population evolution shows a clear preference of the system towards $2\omega_1$ (with $P^{\rm max}_{2\omega_1} \approx 4.5\times P^{\rm max}_{\omega_4}$), which was not visible by analyzing the IR spectra only.

It should be emphasized, that the obtained population evolution is fundamentally different from what would be obtained using the Markovian approximation. Within this approximation, an exponential decay of the population in $v = 1$ would be obtained, and no oscillations or recurrences would be seen.\cite{Morse_Surface_2012,bouakline_quantum-mechanical_2019,fischer_non-markovian_2022} As shown in our previous study,\cite{Attal_EBS_model_2024} a non-Markovian method such as the time-convolutionless approach\cite{Breuer:2001aa} that still traces out the bath, would be able to see the small oscillations in the dynamics but would not reproduce the recurrences observed here. Only a non-Markovian method with an explicit representation of the bath seems to reproduce such features.

\paragraph{Temperature effects}

\begin{figure}[t]

		 \centering    
   \includegraphics[width=1\linewidth]{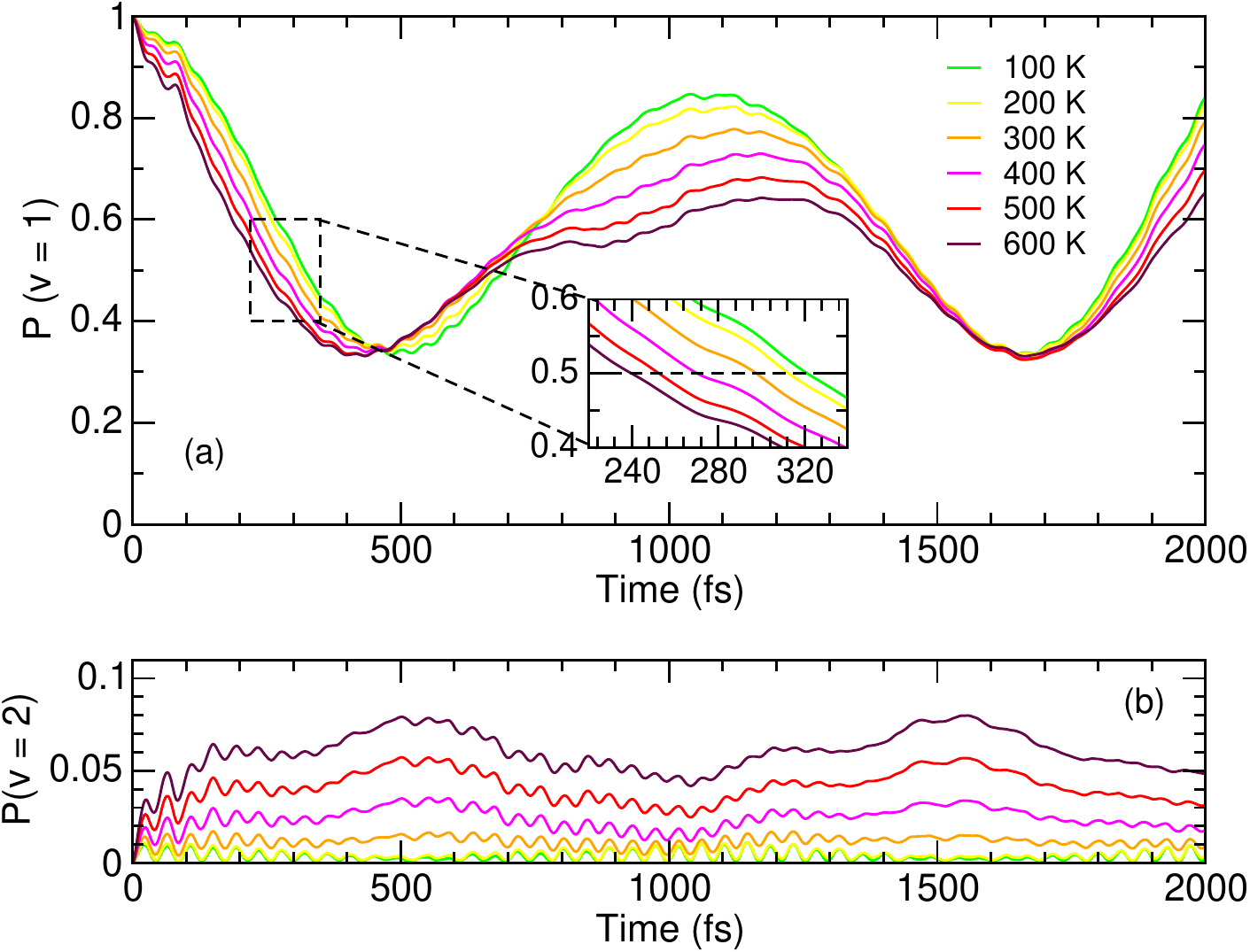}
		   \caption[Time-evolution of the population in $v = 1$ and $v = 2$ at different temperatures] {Time-evolution of the population in (a) the first excited state of the mode of interest, and (b) its second excited state, for bath temperatures ranging from 100~K to 600~K. In panel (a), the inset emphasizes the half-life times obtained at different temperatures, i.e., the time at which the population in $v = 1$ drops below 50\%.}
        \label{fig:10modes_IVR_T}
\end{figure}

In order to reproduce the time evolution of the populations at non-zero temperatures, we have performed a large number of calculations with different initial bath energies. This procedure allows us to reproduce the behavior of a bath at a given non-zero temperature. The system always starts in $v=1$, as if it were prepared by a laser pulse, for example.
We have computed more than 100 trajectories with initial bath energies ranging from 0 to 3100~cm$^{-1}$. With this energy range, it was possible to obtain IVR of the model system for temperatures up to 600~K. The time evolution of the populations in $v = 1$ and $v = 2$ is shown in Fig.~\ref{fig:10modes_IVR_T}, for bath temperatures $T = 100$ -- 600~K. A significant temperature effect can be seen in their evolution, with a faster relaxation of the first excited state when the temperature increases. 
As emphasized in the inset of Fig.~\ref{fig:10modes_IVR_T}(a), the half-life time of the system (i.e. the time needed to empty half of the initial state $v=1$) drops by 25\% on the considered energy range (from 322~fs for 100~K to 240~fs at 600~K). The faster decrease of $P(v = 1)$  comes from the possibility for the mode of interest to reach higher excited states ($v > 1$)  when the bath contains enough energy. In that case, the population in $v = 1$ can either relax to the ground state by giving its energy to the bath, or be excited to $v > 1$ by taking some energy from the thermally excited bath. This second pathway also takes population from the initial state, leading to a faster decrease of $P(v = 1)$. As can be seen Fig.~\ref{fig:10modes_IVR_T}(b), it also leads to a significant increase in the population of $v=2$, with its maximum value rising from less than 1\% to almost 8\%.
Note that $P(v = 1)$ reaches the same minimal value ($\sim 35$\%) for all considered temperatures, since the population transfer efficiency is limited by resonance conditions rather than by temperature effects.

The temperature dependence of the population evolution is even more visible at longer times, where the partial recurrence that brought back 85\% of the population in $v = 1$ in Fig.~\ref{fig:IVR_10modes_v1_m0} drops by 20\% between 100~K and 600~K. The progressive loss of the recurrence structure is due to two factors. First, a part of the population goes to the second excited state instead of coming back to $v =1$ (see Fig.~\ref{fig:10modes_IVR_T}(b)).  Second, more and more bath modes are involved in the dynamics when the temperature increases and the wavepacket gets diluted inside the bath. Thus, it is less likely to reform in its initial state. 

These first results are very encouraging for the possibility of using the EBS method to compute vibrational spectra and internal dynamics of complex molecules, even at high temperatures.  
With the information provided by EBS calculations, the spectra can be assigned, and the internal population transfers can be analyzed. This first application has shown the importance of including quartic and bi-linear coupling terms in the bath coordinates when considering intra-molecular applications. It has also shown that the EBS method is able to deal with such coupling terms and to reproduce full-dimensional calculations. Still, the model system above is rather small, and the effective states involved have very low DOSs. This is a quite favorable regime, where the EBS method tends toward exact treatment. To further test our method, we turn to a larger and more realistic system, namely the phenylacetylene molecule (Ph-Ace).


\section{Application to Phenylacetylene}
\label{sec:Ph-Ace}
\subsection{Context}

Following its recent detection in the interstellar medium,\cite{Loru_Ph_Ace_2023} phenylacetylene (C$_8$H$_6$) has attracted a lot of attention, and several groups have studied its gas-phase infrared spectroscopy.\cite{lacinbala_aromatic_2022,Singh_Ph_Ace_2023, Esposito_Ph_Ace_2024, Esposito_Ph_Ace_Acetylenic_2024}  Their studies, both theoretical and experimental, show that Ph-Ace presents interesting anharmonic features in its aromatic and acetylenic regions, with a very rich spectrum due to large anharmonicities and internal couplings between its vibrational modes.
In this section, we compute the finite-temperature infrared absorption spectrum of Ph-Ace, and compare the EBS results with experimental data and with calculations from the literature. Population transfers inside the molecule are also investigated on one example.

\subsection{Calculation details}
The PES used to model Ph-Ace was obtained using density functional theory (DFT) with the B97-1 functional and the TZ2P basis set. The $g = 36$ normal modes of Ph-Ace were extracted from this potential. The associated harmonic frequencies range from 138~cm$^{-1}$ to 3455~cm$^{-1}$ and are numbered by increasing frequency (see App.~\ref{App6B}). To match the EBS Hamiltonian, the DFT-based PES was fitted into a quartic expansion such as the one presented in Sec.~\ref{subsec:S-B_Hamiltonian}. Inspired by Ref.~\onlinecite{Esposito_Ph_Ace_2024}, we focus on the mid- to far-infrared region of the Ph-Ace spectrum.
Each of the twenty-three IR active normal modes having a frequency between 100 and 1700~cm$^{-1}$ was successively taken as the mode of interest of the EBS method, and its spectrum was computed at 300~K. 
The twenty-three partial spectra were then added to one another and compared with other experimental and theoretical spectra. To obtain relevant intensities, each EBS partial spectrum is scaled by the square of its transition dipole $|\partial \mu/ \partial Q_{i_0}|^2$ (see App.~\ref{App6B}). Absorption spectra at 300~K were computed using  EBS parameters $N_v = 5$, $M = 10\,000$, and $\Delta E = 1$~cm$^{-1}$ and convoluted with a Gaussian having an FWHM of 5~cm$^{-1}$.

\subsection{Infrared absorption spectrum}
\begin{figure}[t]
    \centering
    \includegraphics[width=1\linewidth]{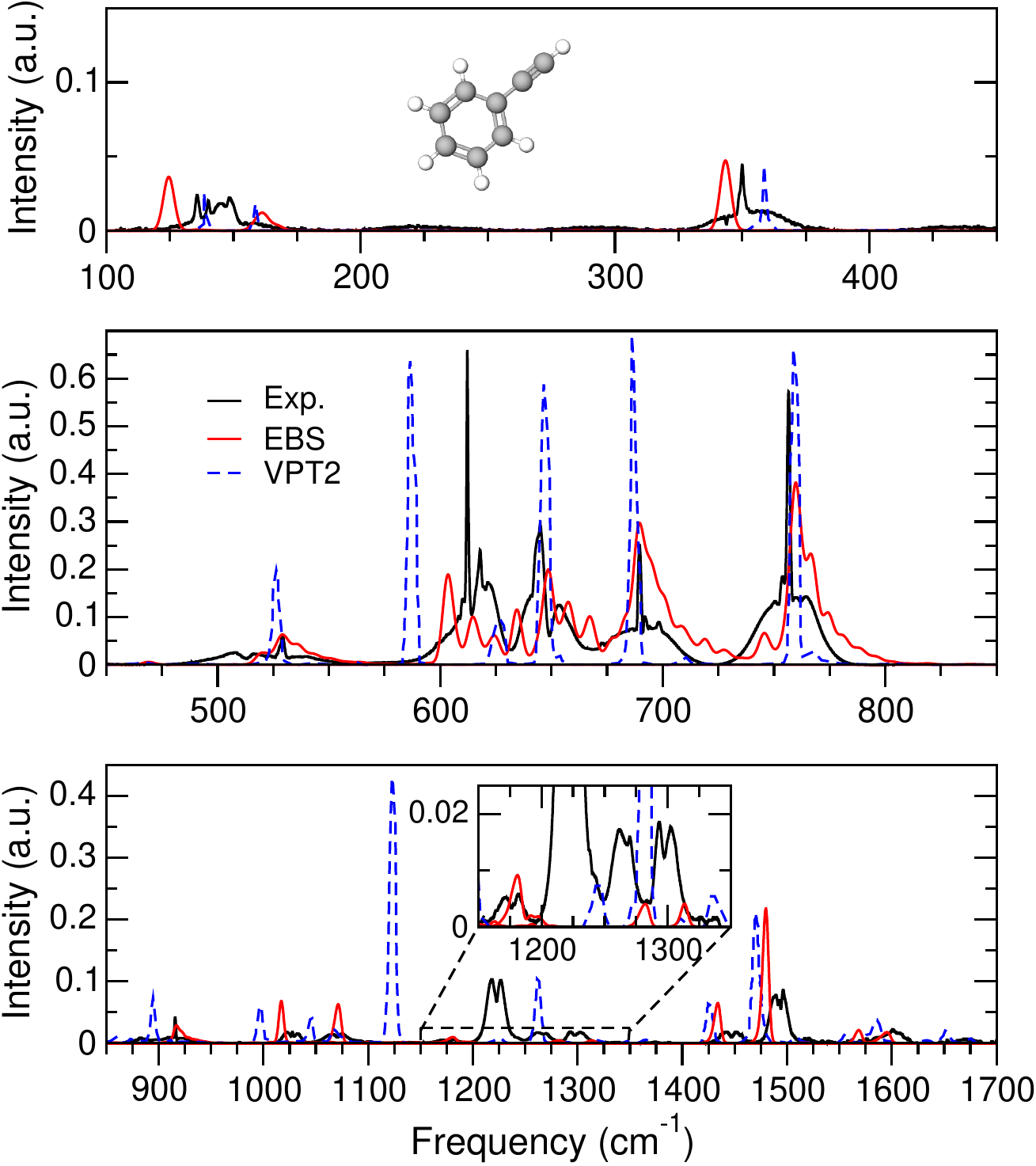}
    \caption{Absorption spectrum of phenylacetylene (molecule shown in the upper panel). Experimental spectrum (solid black line) and VPT2-based theoretical calculations\cite{Esposito_Ph_Ace_2024} (dashed blue line) are compared with the EBS spectrum computed at 300~K and red-shifted by 20~cm$^{-1}$ (solid red line). The inset in the lower panel shows the small EBS features in the 1200 -- 1300~cm$^{-1}$ region.  Intensities are in arbitrary units; note the different scales between the three panels.}
    \label{fig:Ph_Ace_Spectrum}
\end{figure}

The EBS infrared absorption spectrum at 300~K was obtained from the above procedure in the 100 -- 1700~cm$^{-1}$ region. Since a systematic shift of about $+20$~cm$^{-1}$ was noticed between EBS and experimental spectra, Fig.~\ref{fig:Ph_Ace_Spectrum} displays the EBS results red-shifted by $20$~cm$^{-1}$. This shift was chosen to better match the experimental data, especially in the crowded 600 -- 800~cm$^{-1}$ region. The experimental spectrum has been recorded by Marie-Aline Martin and Olivier Pirali using the absorption spectrometer described in Ref.~\onlinecite{Pirali2013}, with an effective absorption length of 140~m and for a pressure of 5~$\mu$bar. Since the EBS calculations yield relative intensities, the comparison with the experimental spectrum is made by setting the intensity of mode 3 (near 350~cm$^{-1}$) to be the same as the experimental one. This mode was chosen as a reference because it is spectrally well isolated. The overall correspondence between our (shifted) calculations and the experimental spectrum is good, EBS bands positions and intensities being coherent with experimental ones. A noteworthy exception appears in the 1200 -- 1300~cm$^{-1}$ region where the EBS intensities are much lower than experimental ones and the band around 1225~cm$^{-1}$ is not reproduced (see inset of Fig.~\ref{fig:Ph_Ace_Spectrum}).

To understand the difference between the EBS and experimental spectra in that region, the EBS results are also compared with theoretical calculations from Ref.~\onlinecite{Esposito_Ph_Ace_2024}, which are based on a VPT2 treatment of a PES obtained at the B3LYP/N07D level. Their calculations take into account anharmonicities and transitions involving up to three quanta (hence, it includes overtones and combination bands). Overall, EBS calculations agree quite well with these results.  Except in the 1200 -- 1300~cm$^{-1}$ region, where VPT2 calculations also seem to struggle to reproduce the main experimental feature near 1225~cm$^{-1}$. However, VPT2 calculations from Ref.~\onlinecite{Esposito_Ph_Ace_2024} obtain an intense band at 1146~cm$^{-1}$, which they associate with that experimental feature. This band is identified by the authors as an overtone of the out-of-plane acetylenic CH bending mode ($\omega_h =$ 610~cm$^{-1}$) that would be strongly shifted from the experimental band due to difficulties in modelling such modes with DFT.\cite{LEE2021,Esposito_Ph_Ace_2024} These difficulties are accentuated for higher-order transitions such as overtones.\cite{FORTENBERRY201513, Esposito_Ph_Ace_2024} 
Due to the linear form of the dipole used in this work, transitions with $\Delta v > 1$ cannot be obtained properly. This would require a non-linear dipole function, which is in principle, possible within the EBS method. Hence, our calculations are currently unable to reproduce the experimental band near 1225~cm$^{-1}$. Note that the experimental spectrum is rotationally resolved and thus presents PQR branches that the EBS or VPT2 calculations are not able to reproduce.

\begin{figure}[t]
	\centering
	\includegraphics[width=1\linewidth]{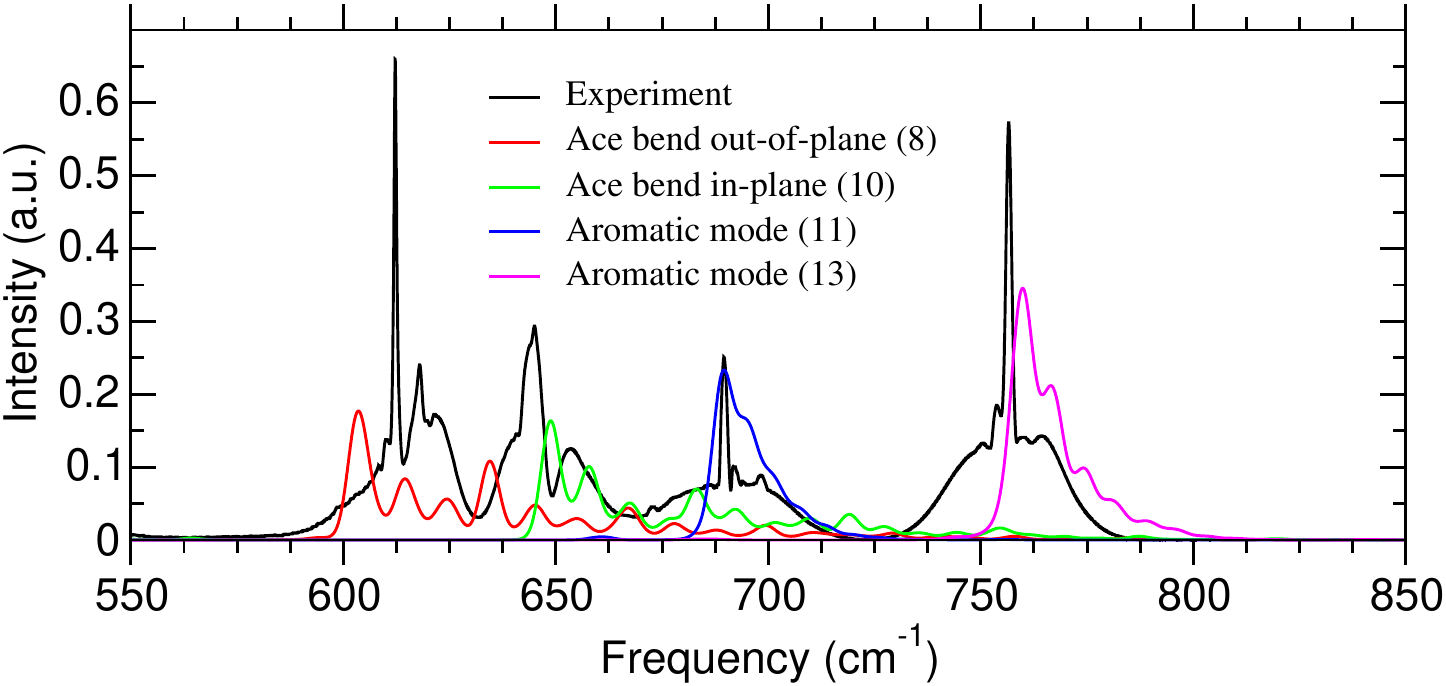}
	\caption{Absorption spectrum of phenylacetylene in the region 550--850~cm$^{-1}$ at 300~K. Relevant (red-shifted) individual partial spectra obtained with the EBS method are compared with the experimental spectrum. Numbers in the legend correspond to the index of the associated normal mode (see App.~\ref{App6B}).}
	\label{fig:Ph_Ace_Spectrum_zoom}
\end{figure}

Figure~\ref{fig:Ph_Ace_Spectrum_zoom} is a focus on the crowded  550--850~cm$^{-1}$ region of the spectrum and shows the partial spectra obtained by the EBS method.
In this figure, each EBS spectrum is the result of a single EBS calculation, taking the indicated mode as the system. The convoluted sum of these partial results provides the spectrum shown in Figure~\ref{fig:Ph_Ace_Spectrum}. Accessing the individual EBS spectra allows for a more detailed analysis of the spectrum since they provide direct information on which normal mode is involved in the experimental bands. Figure~\ref{fig:Ph_Ace_Spectrum_zoom} also shows that individual EBS spectra display complex structures due to anharmonicities and temperature effects. 

\subsection{Intramolecular vibrational redistribution}

\begin{figure}[t]
    \centering
    \includegraphics[width=1\linewidth]{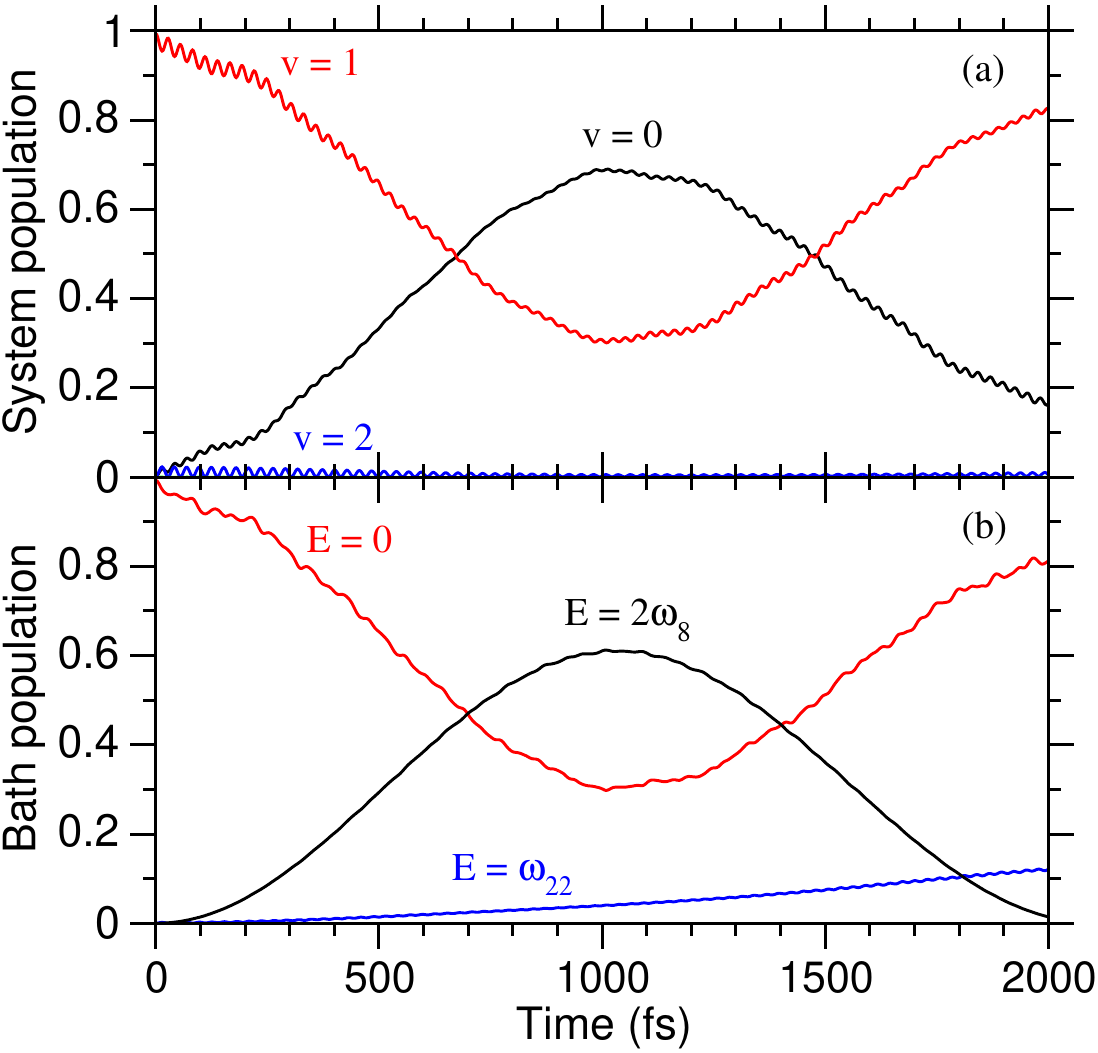}
    \caption{Time-evolution of the system and bath populations when taking mode 23 as the mode of interest, and starting from $\ket{v = 1, m = 0}$. (a) Evolution of the population in the vibrational states of the mode of interest. (b) Evolution of the population in the effective bath states, labeled by their energy. The bath is made of all the normal modes of Ph-Ace except mode 23.}
    \label{fig:Mode_23_29_IVR}
\end{figure}

We now investigate the possibility of computing IVR in such a realistic system on the example of mode 23 ($\omega_{23} = 1217$~cm$^{-1}$). This aromatic C-H bending mode is seen as the system in the EBS method, and the rest of the normal modes are included in the bath. The evolutions of the system and bath populations are obtained using the same parameters as before, starting from the system's first vibrational excited state $v = 1$ and from the bath ground state $m = 0$. 
As seen in Fig.~\ref{fig:Mode_23_29_IVR}, mode 23 strongly interacts with other normal modes and loses around 70\% of its population in 1~ps, before a partial recurrence occurs. It mostly interacts with the nearest bath mode in frequency $\omega_{22}=1197$~cm$^{-1}$, and the first overtone of $\omega_{8}=610$~cm$^{-1}$, which displays a quasi-Fermi resonance with mode 23. 
Normal mode analysis indicates that mode 22 is another aromatic C-H bending mode, whereas mode 8 corresponds to the out-of-plane bending of the acetylenic C-H.\cite{Singh_Ph_Ace_2023} 
Fig.~\ref{fig:Mode_23_29_IVR}(b) shows that the effective bath state associated with the first overtone of mode 8 gains much more population from its interaction with mode 23 than the one associated with mode 22. This can seem surprising, since mode 8 corresponds to an acetylenic excitation, and one could expect that aromatic modes would interact mostly between themselves. This preference seems to rather come from (quasi-)resonance conditions, since

\begin{equation}
    |\omega_{23} - 2\omega_{8}| = 3~\text{cm}^{-1} <\; |\omega_{23} - \omega_{22}| = 20~\text{cm}^{-1}.
\end{equation}
This example thus illustrates the importance of resonant conditions, which can drive the dynamics even when intuitions based on the spatial localization of the modes would suggest otherwise. It is coherent with previous studies: for example, it was found in Ref.~\onlinecite{Singh_Ph_Ace_2023} that the aromatic modes of Ph-Ace have a strong influence on the spectral response in the region of the acetylenic C-H stretching.


\section{Conclusion and perspectives}
\label{sec:conclusion}
The EBS method is a system-bath approach based on the coarse-graining of the bath modes into effective energy states. This method aims at modelling the quantum vibrational dynamics of one-dimensional systems in contact with a large but finite environment. In a previous article,\cite{Attal_EBS_model_2024} we had presented this new method in the context of an effective vibrational mode interacting with an Ohmic bath, with couplings that were linear in the bath coordinates. In that case, an excellent agreement was found with full-dimensional MCTDH calculations from the literature,\cite{Morse_Surface_2012} but at a much lower computational cost.\cite{Attal_EBS_model_2024} 

In the present contribution, we have extended the EBS method to polynomial couplings in the bath coordinates and adapted its formalism to make it suitable for intra-molecular applications. The idea is to divide the studied molecule into a mode of interest (the system) and an internal bath made of the remaining vibrational modes of the molecule. This intrinsically finite bath cannot be treated using usual open quantum methods since they assume the bath to be infinite.\cite{Breuer:2007ab,Tanimura_review_2020_jcp} Because of that, these methods cannot model an out-of-equilibrium bath, or reproduce the recurrences observed in molecular systems.\cite{Finite_bath_2022} The EBS method aims at situations where such a finite bath is too large to be treated with a full-dimensional method, but where the finite-dimensional effects still play a role, as is the case for large molecules.

In this article, we also detailed how our formalism can provide the spectral response of the mode of interest at finite temperatures. Since the EBS method separates the mode of interest from the other modes, and makes additional approximations on the latter, a given EBS calculation does not provide the full spectrum of the studied molecule but rather a partial spectrum representing the spectral signature of the mode of interest, assisted by the rest of the molecule. It is then possible to compute different partial spectra by changing the mode of interest. These spectra can be added to one another in order to cover a large spectral range. Yet, especially in crowded spectral regions, the sum of the partial spectra is not equivalent to the spectrum that a full-dimensional calculation would obtain. This is hence another approximation introduced by our method.

After discussing the extension of the EBS formalism to polynomial couplings and spectroscopic applications, the method was validated using a 10-mode model system. The EBS absorption spectra were found to agree very well with full-dimensional calculations, even at high temperatures. It was also possible to analyze the EBS spectra and assign transitions. Population evolutions after excitation of the mode of interest have also been obtained. This model system has shown (i) the ability of the EBS method to satisfactorily reproduce IR absorption spectra, including temperature effects, (ii) its ability to compute internal population transfers at finite temperature, and (iii) the importance of including quartic coupling terms in the bath coordinates for molecular applications.

Finally, the phenylacetylene molecule was taken as a realistic example, and EBS absorption spectra at room temperature were successfully compared with experimental and theoretical spectra in the mid- to far-infrared region. However, a more realistic form of the dipole moment seems to be needed to reproduce some overtone and combination bands that are important in the spectrum. Internal population transfers after excitation of a specific mode have also been obtained. Their analysis has shown that resonant or quasi-resonant conditions between different modes can create strong couplings between acetylenic and aromatic modes of the molecule.

Our study of phenylacetylene did not cover the highest frequency modes of the molecule, especially the acetylenic C-H stretching mode, which has been shown to have interesting intra- and inter-mode anharmonic properties.\cite{lacinbala_aromatic_2022,Singh_Ph_Ace_2023, Esposito_Ph_Ace_Acetylenic_2024}
This is due to the very strong coupling between this mode and the in-plane acetylenic C-H  bending mode, and to the strong anharmonicity of both modes. This makes it difficult to assume one of them to be harmonic or to correctly compute their couplings within second-order perturbation theory.\cite{Loise_these_2024} A natural extension of this work would be to extend the EBS method from a one-dimensional to a two-dimensional system to correctly treat the anharmonicities and strong coupling occurring in the acetylenic end of the molecule. 

Another perspective would be towards emission spectroscopy. This is particularly interesting for molecules of astrophysical interest, such as polycyclic aromatic hydrocarbons which have been suggested to be at the origin of the so-called aromatic infrared bands.\cite{Leger_PAH_1984, Sellgren8PAH_1984,Allamandola_PAH_1985} Coefficients for spontaneous emission can be deduced from absorption coefficients, and emission spectra can then be obtained using a density matrix approach.\cite{cohen2012processus,Grynberg_Aspect_Fabre_book} Since the EBS method can provide absorption coefficients, it would be possible to compute emission coefficients and to obtain time-resolved emission spectra using the Liouville-von Neumann equation.\cite{Breuer:2007ab}

\begin{acknowledgments}
The authors thank Marie-Aline Martin and Olivier Pirali for providing the experimental spectrum used in Fig.~\ref{fig:Ph_Ace_Spectrum} and \ref{fig:Ph_Ace_Spectrum_zoom}. 
L.A. acknowledges financial support from MESRI through a PhD fellowship granted by the EDOM doctoral school.
\end{acknowledgments}

\section*{Author Contributions}
\textbf{Lo\"ise Attal: }
Conceptualization (lead),
Investigation (lead),
Methodology (lead),
Software (lead),
Visualization (lead),
Writing -- Original Draft (lead),
Writing -- Review \& Editing (equal);
\textbf{Cyril Falvo: }
Conceptualization (supporting),
Investigation (supporting),
Methodology (supporting),
Software (supporting),
Writing -- Review \& Editing (equal);
\textbf{Pascal Parneix:} 
Conceptualization (lead),
Investigation (supporting),
Methodology (supporting),
Software (supporting),
Supervision (lead),
Writing -- Review \& Editing (equal).

\section*{Conflicts of interest}
There are no conflicts to declare.

\section*{Data Availability Statement}
The data that support the findings of this study are available
from the corresponding author upon reasonable request.

\appendix

\section{Details on difference bands calculations}
\label{AppA}

Difference transitions are either associated with operator $\hat{Q}_{j}^{+}\hat{Q}_{k}^{-}$ and are of the form $ (n_j, n_k) \to (n_j+ 1, n_k-1)$, or to $\hat{Q}_{j}^{-}\hat{Q}_{k}^{+}$ leading to transitions $ (n_j, n_k) \to (n_j -1, n_k+1)$. 
In both cases, they are characterized by the transition parameter 
$\Delta m^-= \left[\frac{\hbar (\omega_j-\omega_k)}{\Delta E} \right] = m_j - m_k$.
Here we focus on the effective expression of $\hat{Q}_{j}^{+}\hat{Q}_{k}^{-}$, and write
\begin{multline}
	\hat{Q}_{j}^{+}\hat{Q}_{k}^{-} = \sum_{m = m_k}^{M -m_j-1}  \sum_{n_j = 0}^{N_j^{\rm (eff)}}\sum_{n_k = 1}^{N_k(m,n_j)}\bra{n_j+1} \hat{Q}_j \ket{n_j}\bra{n_k-1} \hat{Q}_k \ket{n_k} \\ \sum_{{\bf n}^* \in m^*}\ket{n_j+1,n_k-1,{\bf n}^*}\bra{n_j,n_k,{\bf n}^*},
\end{multline}
where ${\bf n}^* = \{n_l\}_{l \ne j,k}$. 
In the expression above, the value of $m$ is constrained by both $n_k \to n_k -1$ and $n_j \to n_j +1$. Since the transition $n_k \to n_k -1$ is only possible if $n_k \geq 1$, the smaller accessible state is constrained by $m \geq m_k$. In the same way, the state containing $n_j+1$ must be defined (i.e. $\leq M-1$), leading to $m \leq M - m_j - 1$.
 
The microstate $(n_j,n_k,{\bf n}^*)\in \ket{m}$ is hence constrained by 
\begin{align}
	\left\{\begin{array}{lll}
		0\leq n_j\leq N_j^{\rm (eff)} \\
		1\leq n_k\leq N_k(m,n_j) \\
		n^* \in m^*_{j,k} \end{array}\right. ,
\end{align}
with $N_j^{\rm (eff)} \ne N_j(m)$ since the condition $n_k \geq 1$ constrains the upper value of $n_j$. The maximum value of $n_j$ in $\ket{m}$ is such that

\begin{align}
	N_j^{\rm (eff)}\times m_j + 1\times m_k = m,
\end{align}
hence 
\begin{align}
	N_j^{\rm (eff)}m_j = m - m_k, 
\end{align}
and $N_j^{\rm (eff)} = N_j(m-m_k)$. The microstate $(n_j,n_k,{\bf n}^*)$ is therefore constrained by 
\begin{align}
	\left\{\begin{array}{lll}
		0\leq n_j\leq N_j(m-m_k)  \\
		1\leq n_k\leq N_k(m,n_j) \\
		n^* \in m^*_{j,k} \end{array}\right. ,
\end{align}
which is equivalent to 
\begin{align}
	\left\{\begin{array}{lll}
		0\leq n_j\leq N_j(m-m_k)  \\
		0\leq n_k-1\leq N_k(m,n_j)-1\\
		n^* \in m^*_{j,k} \end{array}\right. .
\end{align}
Furthermore, $N_k(m,n_j)$ is define by 
\begin{align}
	N_k(m,n_j)m_k + n_jm_j = m,
\end{align}
hence 
\begin{align}
	\left(N_k(m,n_j)-1\right)m_k + n_jm_j = m-m_k, 
\end{align}
and $N_k(m,n_j)-1 = N_k(m-m_k,n_j)$. Therefore we have
\begin{align}
	\left\{\begin{array}{lll}
		0\leq n_j\leq N_j(m-m_k)  \\
		0\leq n_k-1\leq N_k(m-m_k,n_j)\\
		n^* \in m^*_{j,k} \end{array}\right. ,
\end{align}
and the microstates involved in the transition are fully characterized by the effective state $\ket{m-m_k}$, and its DOS $\rho(m-m_k)$ must be used to define the effective operator $\hat{Q}_{j}^{+}\hat{Q}_{k}^{-}$. 
As in Ref.~\onlinecite{Attal_EBS_model_2024}, this leads to the following transformations:
\begin{align}
	\ket{n_j,n_k,{\bf n}^*} &\to \frac{1}{\sqrt{\rho(m-m_k)\Delta E}} \ket{m}, \\
	\ket{n_j+1,n_k-1,{\bf n}^*} &\to \frac{1}{\sqrt{\rho(m-m_k) \Delta E}} \ket{m+ \Delta m^-}.
\end{align}
From this we write the effective representation of $\hat{Q}_{j}^{+}\hat{Q}_{k}^{-}$ as 
\begin{align}
	\hat{Q}_{j}^{+}\hat{Q}_{k}^{-} = & \sum_{m = m_k}^{M -m_j-1}  \sum_{n_j = 0}^{N_j(m-m_k)} \bra{n_j+1} \hat{Q}_j \ket{n_j}\nonumber \\ 
	& \times \sum_{n_k = 1}^{N_k(m,n_j)}  \bra{n_k-1} \hat{Q}_k \ket{n_k} \\ 
	& \times\sum_{{\bf n}^* \in m^*}\frac{1}{\rho(m-m_k) \Delta E}\ket{m+\Delta m^-}\bra{m}. \nonumber
\end{align}

The sum $\sum_{{\bf n}^* \in m^*}$ contains $\rho^{(j,k)}(m-m_k)\Delta E$ microstates and we hence have

\begin{align}
	\sum_{{\bf n}^* \in m^*}\frac{1}{\rho(m-m_k) \Delta E}  &= \frac{\rho^{(j,k)}(m-m_k)\Delta E}{\rho(m-m_k) \Delta E} \nonumber \\
	 &=  \mathbb{P}(m-m_k,n_j,n_k-1)\, ,
\end{align}
which is the probability to have $n'_j = n_j$ and $n'_k = n_k-1$ in the effective state $\ket{m-m_k}$. 
Hence, we finally obtain
\begin{align}
	\hat{Q}_{j}^{+}\hat{Q}_{k}^{-} = & \sum_{m = m_k}^{M -m_j-1}  \sum_{n_j = 0}^{N_j(m-m_k)}\bra{n_j+1} \hat{Q}_j \ket{n_j}\nonumber \\ 
	& \times \sum_{n_k = 1}^{N_k(m,n_j)}\bra{n_k-1} \hat{Q}_k \ket{n_k} \\ &\times\mathbb{P}(m-m_k,n_j,n_k-1)\ket{m+\Delta m^-}\bra{m}. \nonumber
\end{align}


\section{Parameters of the 10-mode model system}
\label{App6}
In this appendix, we provide the harmonic frequencies and anharmonic parameters of the PES of the 10-mode model system used in Sec.~\ref{sec:Spectra_10modes}.
The selected mode of interest is $i_0= 3$. In the EBS method, only $i_0$ has intra-mode anharmonicity, and only the coupling terms connecting $i_0$ to the bath modes are considered.

\begin{table}[!h]
    \centering
    \begin{tabular}{p{0.2\linewidth}|p{0.2\linewidth}|p{0.2\linewidth}|p{0.2\linewidth}}
    \hline\hline
    i & $\omega_i$ & $\alpha_{iii}$ & $\beta_{iiii}$ \\ \hline
 1 & 410  &  0  &       0  \\
2& 560   &  0  &        0   \\
3 &800 &   0.0024    &    0.0002 \\
4&830 &    0 &         0   \\
5 &1260  & 0   &      0     \\
6& 1450 & 0    &      0   \\
7& 1510  &  0   &       0  \\
8& 1660  &   0   &       0 \\
9& 1712  &  0   &       0  \\
 10&1860  &  0    &      0  \\
 \hline\hline
    \end{tabular}
    \caption{Harmonic frequencies (in cm$^{-1}$) and intra-mode anharmonicities (in atomic units) for the ten modes of the model system.}
\end{table}
 \begin{table}[!h]
    \centering
    \begin{tabular}{p{0.2\linewidth}|p{0.2\linewidth}|p{0.2\linewidth}|p{0.2\linewidth}}
    \hline\hline        
    k & $\alpha_{i_0i_0k}$ & $\alpha_{i_0kk}$& $\beta_{i_0i_0kk}$ \\ \hline
    1  & -0.0014 & 0.0009& 0.00015\\
    2 &  0.0008 &  0.0013&0.0005\\
    4   &  -0.0018& 0.0006& 0.0025\\
    5  &-0.0006 &-0.0010& -0.0001\\
    6 &  0.0008 &  0.0012&0.00325\\
    7  & 0.0022& 0.0014 & -0.0010\\
    8  &  -0.0026  &   0.0020 &0.00035\\
    9 & 0.0004 & 0& 0.00055\\ 
     10  &0.0050 &  -0.0003& 0.0003\\
 \hline\hline
    \end{tabular}
    \caption{Two-mode coupling coefficients (in atomic units) associated with coupling terms $Q_{i_0}^2Q_k$, $Q_{i_0}Q_k^2$, and $Q_{i_0}^2Q_k^2$, respectively. With $i_0 = 3$ the mode of interest and $k \ne i_0$ a  bath mode.}
\end{table}

\begin{table}[!h]
    \centering
    \begin{tabular}{p{0.1\linewidth}|p{0.1\linewidth}|p{0.2\linewidth}||p{0.1\linewidth}|p{0.1\linewidth}|p{0.2\linewidth}}
    \hline\hline        
    j & k&  $\alpha_{i_0jk}$ &j & k&  $\alpha_{i_0jk}$ \\ \hline
 1 &   2 & 0  &  4 &  8 & 0.00008  \\    
 1  &  4 & 0.004   & 4  &  9 & 0.00026      \\ 
  1  &  5 & 0   &  4 &  10 & -0.0030 \\
  1  &  6&  0     &    5   & 6 & 0.00004  \\
  1  &  7&  0.00012 &  5   & 7 & 0  \\  
  1  &  8& 0    &5  &  8 & 0   \\
  1  &  9 & 0.0002  & 5  &  9 & 0.0002  \\
  1  &  10 & 0.0010  &5  &  10 & 0    \\ 
  2  &  4 & 0   &6  &  7 & 0.00010 \\
  2  &  5 & 0.00046 &  6  & 8 & -0.0014 \\   
  2  &  6 & 0 & 6  &  9 & 0 \\
  2  &  7 & 0.0002 & 6  &  10  &0.00016 \\
  2  &  8 & 0  & 7  &  8 & 0.00008 \\ 
  2  &  9 & 0  & 7  &  9  &0  \\ 
  2 &  10 & -0.0016 &7  &  10 & -0.0004 \\  
  4  &  5 & 0.00048&8  &  9 & 0.00004 \\   
  4  &  6 & 0   &  8  &  10 & -0.00090 \\
  4  &  7& 0  &    9  &  10 & 0.00024 \\   
 \hline\hline
    \end{tabular}
    \caption{Three-mode coupling coefficients (in atomic units) associated with coupling terms $Q_{i_0}Q_jQ_k$, with $i_0 = 3$ the mode of interest and $j,k \ne i_0$ two different bath modes ($j \ne k$). Since $\alpha_{i_0jk} = \alpha_{i_0kj}$, the coefficients are only given for $k >j$. }
\end{table}

\eject 
\section{Harmonic frequencies and intensities of phenylacetylene}
\label{App6B}

\begin{table}[!h]
    \centering
    \begin{tabular}{c|c|c||c|c|c}
    \hline\hline        
    Mode & Freq.  & IR intensity &Mode & Freq. & IR intensity\\ \hline
1& 138.5& 0.179 &{19}& {1044.5}&0.051\\     
2& 157.5& 0.104 &{20}& {1096.8} &0.060\\ 
3& 361.0 & 0.125 &{21}& 1179.9 &0.000 \\
4& 407.0 & 0.000 &{22} &1196.6 &0.000\\
5& 469.3 & 0.007 &{23}& 1215.5 & 0.014\\ 
6& 531.0 & 0.099 &{24}& 1308.5 &0.004\\  
7& 543.2 & 0.179  &{25}& 1349.7 &0.003  \\
8& 610.7& 1.000 & {26} &1470.2&0.043\\
9& 631.7  & 0.023 & {27}& 1518.0&0.131 \\
10& 664.9 & 0.856 &{28} &1604.5&0.013\\
11& 703.2 & 0.619 & {29}&{1633.7}&0.016 \\ 
12& 771.0  & 0.048 &30& 2186.6 &0.055\\
13& 772.2  & 0.881 & 31& 3164.0&0.001\\ 
14& 855.1& 0.000 &32& 3173.0 & 0.013\\ 
15& 934.5 & 0.059 &  33 &3182.9&0.035\\
16& 987.7 &0.000 &34& 3190.3&0.040\\
17& 1003.4& 0.001 & 35 &3194.1&0.013\\
18& 1011.6 &0.000 &36& 3455.0 &0.369\\   
 \hline\hline
    \end{tabular}
    \caption{Harmonic frequencies of the 36 normal modes of phenylacetylene obtained by DFT using the B97-1/TZ2P basis set. The modulus square of the transition dipole along each coordinate is also given (IR intensity). The intensities have been re-normalized so that the maximum value is equal to 1. Only re-normalized intensities above 0.001 are considered in the calculations. }
\end{table}

\eject 

\section*{References}
\bibliography{bibliography}

\end{document}